\documentclass{jaa}
\usepackage{natbib}
\usepackage{hyperref}
\hypersetup{colorlinks = true,
	allcolors  = black, 
	citecolor  = blue,
	anchorcolor = blue}
\makeatletter
\def\p@figure{\color{blue}}

\usepackage{graphicx}

\begin{document}\sloppy
	
	\title{Upcoming SKA precursor surveys and sensitivity to HI mass function}
	
	\author{Sauraj Bharti and J S Bagla}
	\affilOne{Indian Institute of Science Education and Research Mohali,
		Knowledge City, Sector 81, SAS Nagar, Punjab 140306,
		India\\}
	
	\twocolumn[{
		
		\maketitle
		
		\corres{ph19054@iisermohali.ac.in}

		
		\begin{abstract}
			We describe a simulation for the distribution of galaxies
			focusing on the atomic Hydrogen content.
			Our aim is to make predictions for surveys of galaxies using the
			redshifted $21$~cm line emission.
			We take the expected distribution of HI masses, circular velocities,
			sizes of galaxies and orientations into account for this
			simulation. 
			We use the sensitivity of ASKAP and MeeKAT radio telescopes to
			estimate the number of detections of HI galaxies in upcoming surveys.
			We validate our simulation with earlier estimates carried out by
			using some of these considerations.  
			We show that unlike earlier simulations that take some of the
			factors into account, the predicted number of galaxies and their
			distribution across masses changes significantly when all of these
			are accounted for.
			We describe our predictions for the MIGHTEE-HI and WALLABY surveys for
			blind detection of galaxies using the redshifted $21$~cm radiation.
			We study the dependence of the predicted number of detections on
			the HI mass function.
			We also describe our future plans for improving the simulation.
		\end{abstract}
		\keywords{Galaxies -- Simulations -- Atomic Hydrogen}
	}]
	
	\artcitid{\#\#\#\#}
	\volnum{000}
	\year{0000}
	\pgrange{1--17}
	\setcounter{page}{1}
	\lp{17}

	\section{Introduction}
	
	The redshifted $21$~cm line is a key tracer of atomic Hydrogen (HI)
	gas in galaxies.
	It offers a unique window on the evolution of galaxies by probing the
	neutral inter-stellar medium (ISM).
	The HI $21$~cm line is very weak and remains hard to detect at higher
	redshift with the limited sensitivity of radio telescopes.
	This is expected to change with the surveys planned using SKA\footnote{\url{https://www.skatelescope.org/the-ska-project/}} precursors. These surveys, expected to be much more sensitive than the last generation surveys like ALFALFA \citep{1, 2} and HIPASS \citep{3}, 
	will probe galaxy populations and the large scale structure in this window out to much larger distances. Earlier surveys have given us adequate information about the HI mass
	function (HIMF) at low redshifts \citep{4} and
	combining with optical observations \citep{5} has already given us a wealth of
	insight into the correlation of colors, optical properties and the HI
	mass \citep{6, 7, 8}, and circular velocities of galaxies \citep{9}. 
	The long term plan is to develop an insight into the relation of HI
	mass and star formation and hence formation and evolution of galaxies.  The HIMF is ideally estimated using blind surveys \citep{10, 11, 12, 13}.
	Till date, surveys using the redshifted $21$~cm radiation have been
	shallow, i.e., probing galaxy populations at very low redshifts.
	Surveys extending to intermediate or high redshifts \citep{14, 15}
	mostly rely on stacking to measure average properties of galaxies and
	have small samples.
	In order to probe HIMF at intermediate redshifts, we require high
	sensitivity and observations of a large region in the sky.
	Such observations, coupled with observations of radio continuum that
	are obtained at the same time, will allow us not only to estimate the
	HIMF but also star formation properties of galaxies \citep {16, 17}.

	The target of surveys with SKA precursors is to observe in excess of
	$10^5$ galaxies going up to and beyond $z = 0.2$ so that there are 
	adequate statistics to address a wide variety \citep {18} of questions.
	We focus on two upcoming surveys in this paper: the  Widefield ASKAP
	L-Band Legacy ALL-sky Blind surveY (WALLABY\footnote{\url{http://wallaby-survey.org/overview/}}: \citep{19} and MeerKAT International Giga Hertz
	Tiered Extragalactic Exploration (MIGHTEE-HI\footnote{\url{http://idia.ac.za/mightee/}}: \citep{20}.
	There are various commensal surveys with combination of area and depth
	that are complementary mutually for instance MIGHTEE-HI and LADUMA. The
	MIGHTEE-HI is a wide survey but not as deep as LADUMA and LADUMA is a deep
	but not as wide as MIGHTEE. 
	We are not making predictions for LADUMA in this work. The complementary aspect of LADUMA is useful as the observed data from LADUMA, will be made available to the MIGHTEE-HI survey for continuum science studies as well.
	Predictions from simulations for SKA precursors are crucial in
	designing the future surveys.
	We can simulate the expectations for a variety of survey strategies
	and optimize the strategy keeping our goals in mind.
	The goals here related to extracting information about the physical
	properties of galaxies and their evolution with epoch. 
	The earlier
	predictions made for MIGHTEE-HI survey in Maddox N. {\em et al.} 2020 are based
	on \citep{21} semi-analytical models.
	The earlier work on WALLABY predictions \citep{22} is based on semi-analytical models applied
	to cosmological N-body simulations \citep{23}.
	The semi-analytical method for MIGHTEE-HI case makes use of cosmic
	evolution of dark matter \citep{24} within
	$\Lambda$CDM cosmology. 
	
	In this paper we introduce our simulation method.
	Unlike earlier work, we do not make use of cosmological N-Body
	simulations or semi-analytic models of galaxy formation. 
	We ignore the clustering of galaxies and focus on number counts instead. This is a valid assumption as our mock survey volume is of the order of $10^6$ Mpc$^3$ (i.e. $100$ Mpc length scale). This is slightly larger than the scale of  homogeneity estimated from observations, e.g.,  \citet{Ntelis, Scrimgeour}.

	In order to validate our simulations, we use three different
	models: (I) This model assumes a fixed size ($D_{HI}$) and
	linewidth ($W_{20}$) of the galaxies in the mock catalogue. In this catalogue, we do not scale the size or the line width with the HI mass.  We also ignore the
	inclination effect and assume that all galaxies are seen edge-on, (II) This model takes mass dependent size and 
	circular velocity but ignores the inclination effect (i.e. all galaxies are assumed to be seen edge-on), (III) In this
	model we assume mass dependent size, circular velocity and random orientations (see table \ref{Tab1}) and velocity dispersion.  
	We use Model I for validation, all the models for illustration
	although our final results are based only on Model III.
	We explore sensitivity to the HI mass function as well. 
	
	This paper is organised as follows: in \S{2}, we describe our
	simulations and some related concepts.
	In \S{3}, we present relevant parameters used in WALLABY and
	MIGHTEE-HI survey.
	In \S{4}, predictions for number counts are presented.
	We also discuss the sensitivity of the HI mass function parameters in
	this section. We also present the number of galaxies detected in the map if the images are made at different angular resolutions.
	We conclude and summarise our work in the \S{5}.
	
	\begin{table}[ht]
		\tabularfont
		\setlength{\tabcolsep}{2pt}
		
		\caption{The simulation models of galaxies. All three models do not take the clustering and redshift evolution of the HI-mass function into account.}\label{Tab1} 
		\begin{tabular}{lccccc}
			\topline
			Model& $D_{HI}$ & $W_{20}$&   inclination\\\midline
			I&  $50$ kpc&   $300$ km/s & edge-on \\
			II&   M$_{HI}$ dependent& M$_{HI}$ dependent & edge-on\\
			III&  M$_{HI}$ dependent& M$_{HI}$ dependent & random orientation& \\
			\hline
		\end{tabular}
		\tablenotes{The dispersion of $10$ km/s in the circular velocity of the galaxy is taken in Model III.  See Eqn.8.}
	\end{table}

	\section{Simulations and Mock Catalogs}
	
	We introduce our simulation in this section.
	Our primary focus is on physical properties of galaxies other than clustering.
	We choose to distribute galaxies randomly in space in our simulations.
	Given the large scale surveys that are of interest to us and the very
	large volume to be sampled, this limitation does not impact our
	predictions.
	Key components of our simulation are as follows.
	\begin{itemize}
		\item
		We assume a cosmological model. In this paper we work with a flat
		$\Lambda$CDM cosmology \{$\Omega_m$, $\Omega_{\Lambda}$, $h$\} =
		\{0.3, 0.7, 0.7\} to calculate redshifts, distances and volume.  
		The comoving distance $D_C$ as a function of redshift is given as: \citep{25} 
		\begin{equation}
			D_C = D_{H_0} \int_0^{z} \frac{dz^{\prime }}{E(z^{\prime })}
		\end{equation}
		where $D_{H_0}$ is Hubble distance and $E(z)$ is function of redshift and other cosmological parameters, for $\Lambda$-CDM Cosmology
		\begin{equation}
			E(z) = \bigg[\Omega_{0,m}(1+z)^3 + \Omega_{0,\Lambda} + \Omega_{0,r}(1+z)^4\bigg]^{1/2}
		\end{equation}
		\item
		For a given telescope field of view, $\omega$ is the solid angle of the observing cone. The volume of observing cone can be estimated
		using: 
		\begin{equation}
			dV_C = \omega D_C^2\;dD_C
		\end{equation}
		We populate galaxies with a uniform comoving number density in the survey volume. 
		We used Eq.(3) to generate redshifts from uniform random relative
		volume for both surveys to a redshift (MIGHTEE: $z=0.4$  and
		WALLABY: $z=0.26$). 
		We use the comoving volume as a function of redshift as the distribution function here after suitable normalization. 
		\item
		The HI mass function is taken as an input.  We work with the HIMF from the ALFALFA survey \citep{26}, \citep{27}.
		The assumed HIMF is used to assign HI masses within the range [$10^{7}$~M$_{\odot}$, $10^{12}$~M$_{\odot}$] to galaxies in the
		simulation. The HI mass function of the local universe is normalized and taken as the probability distribution function. The HIMF is defined as the number density of the HI galaxies in logarithmic HI mass bin. The HI mass function $\phi(M_{HI})$  can be expressed as:
		\begin{equation}
			\phi(M_{HI}) = \frac{dn}{d\log(M_{HI})}
		\end{equation}
		where $dn$ is the number density of the objects having a mass within the range [$\log(M_{HI}),\log(M_{HI})+d\log(M_{HI})$]. The HI mass function is well fitted \citep{3}
		by a Schechter function:
		\begin{equation}
			\phi(M_{HI}) = ln(10)\,\phi^* \bigg(\frac{M_{HI}}{M_{HI}^*}\bigg)^{\alpha + 1} \exp{ \bigg(\frac{-M_{HI}}{M_{HI}^*}\bigg)}
		\end{equation}
		where the parameters are $\alpha$: low mass end power law index, $M^*_{HI}$: the knee mass and $\phi^*$ is the normalisation constant.  We also study sensitivity of our predictions to the different assumed HI mass functions. Given that the mass function may evolve in surveys that are sensitive enough to detect individual galaxies out to intermediate redshifts, an evolving mass function can also be studied in principle.  However, we are not considering the evolution in HIMF in this paper.  We consider variations in the prediction with parameters of the HIMF.
		\item
		Circular velocities of galaxies are related to the baryonic mass as follows: \citep{29}
		\begin{equation}
			M_b = A\,\left(\frac{V_f}{100 \mathrm{km/s}}\right)^4
		\end{equation}
		where $A = 4.7\pm 0.6\, \times 10^9 \, M_{\odot}$ and $V_f$ is taken from the outer flat part of the rotation curve of the galaxy. This essentially follows
		from the baryonic Tully-Fisher relation. The baryonic mass $M_b = M_*+1.4M_{HI}$ is the sum of components: neutral HI gas and stellar mass $M_*$, and the molecular gas mass is ignored for the HI-dominated galaxies. The molecular gas shares $<5\%$ of the total baryonic mass, and it is less than the uncertainty in measuring the HI gas content in most cases, therefore we choose to neglect it in the present study. We take $M_{HI}-M_*$ relation from \citep{30}.
		\begin{equation}
			\log M_{HI}= 0.35(\log M_*-10)+9.45
		\end{equation}
		We also
		incorporate a velocity dispersion (assumed to be
		$10$~km/s) in our model of galaxies. This becomes relevant when we consider
		inclinations of galaxies as face-on galaxies can end up with an arbitrarily small line width without the dispersion. This does not impact the line width in most other cases.  
		
		We take a fixed velocity width, $W_{20} = 300$km/s for Model I, a mass-dependent velocity for Model II and a mass-dependent velocity with random orientations to account for the added inclination in Model III. 
		
		The inclination-corrected circular velocity projected along the line of sight when combined with the velocity dispersion  gives the linewidth \citep{31}. We assume that galaxies are randomly distributed with respect to the line of sight ($\cos{i}$ is uniformly distributed in [0,1]).  This is used to compute the linewidth of the galaxy in the mock catalog.
		This aspect is taken
		into account only in Model III.  Linewidths are computed using the
		circular velocity, the velocity dispersion, and the inclination angle. The linewidth is denoted by $W_{20}$ ,which is measured at the $20\%$ level of each of the two horns. The linewidth of the galaxy in the mock catalog can be given as follows:
		\begin{equation}
			W_{20} = 2\,\sqrt{V_f^2 \,\sin^2{i} + \sigma_v^2}
		\end{equation}
		Here, $\sigma_v$ is the velocity dispersion and we take it to be $10$~km/s in Model III.  
		Our third model is much more realistic with inclination added, and as we shall see, it predicts the highest number counts at low redshifts compared to Model I and Model II. A detailed analysis of these predictions is given in \S 4.
		
		The velocity width is used to compute the frequency width of the redshifted $21$ cm line with the rest frame frequency, $\nu_{HI}=1420$ MHz. The observed frequency width $\Delta \nu_{ch}$  (spanned across the channels) caused by galaxy rotation can be expressed as
		\begin{equation}
			\Delta \nu_{ch} = \frac{W_{20}}{c}\frac{\nu_{HI}}{(1+z)}
		\end{equation}
		The spectral line from a galaxy typically spans across many channels.
		
		\item
		We use the
		mass-diameter relation \citep{32}  for HI disks to estimate the size of a galaxy. 
		\begin{equation}
			\log\left(\frac{D_{HI}}{\mathrm{kpc}}\right) = 0.51\,\log\left(\frac{M_{HI}}{M_{\odot}}\right)-3.32
		\end{equation}
		In Model I, a fixed size $D_{HI}=50$ kpc is assumed for each galaxy and we use the above relation to estimate sizes of  galaxies in Model II and III.
		\begin{figure*}
			
			\centering\includegraphics[width=.3\textheight]{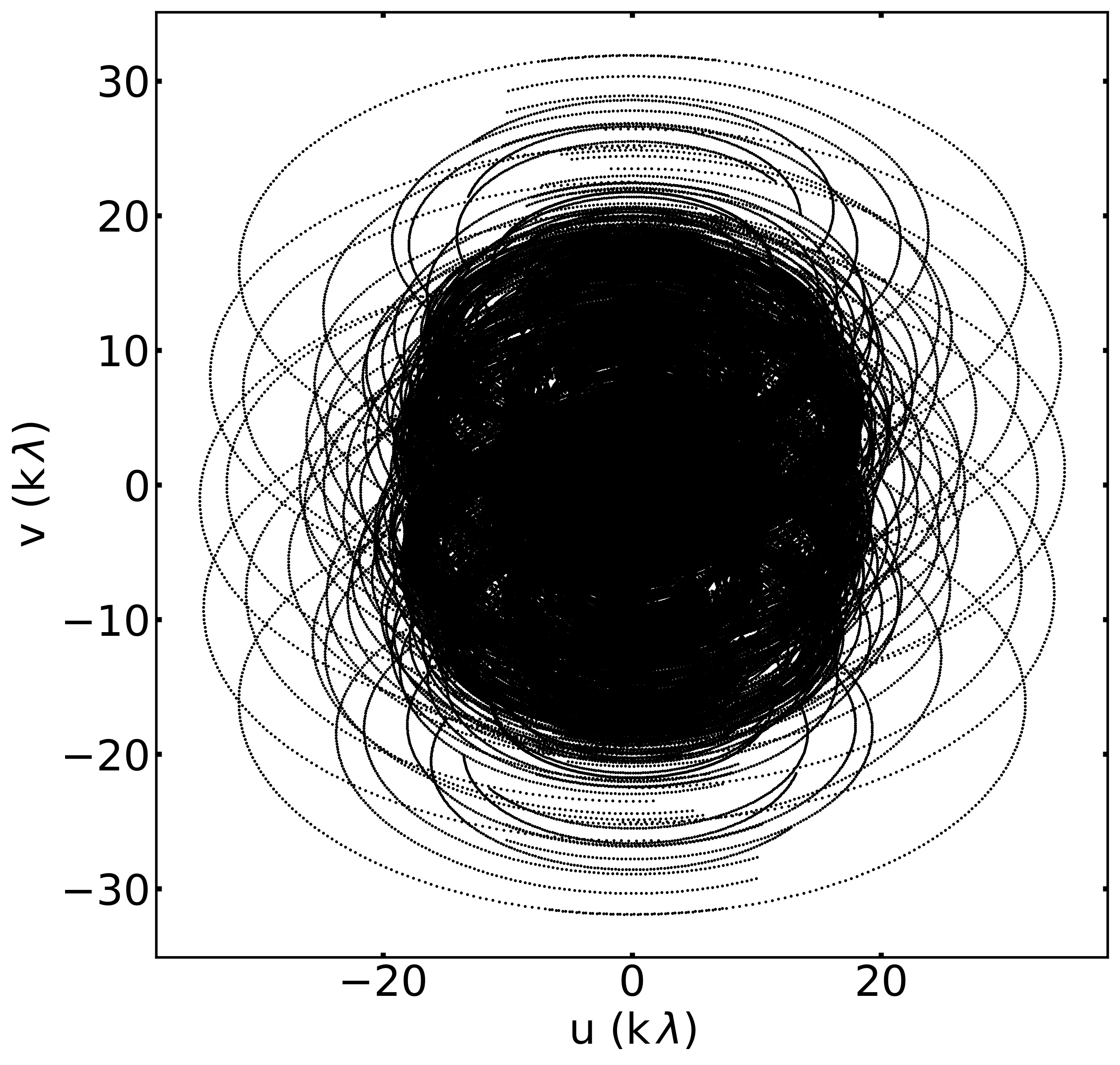}
			\centering\includegraphics[width=.3\textheight]{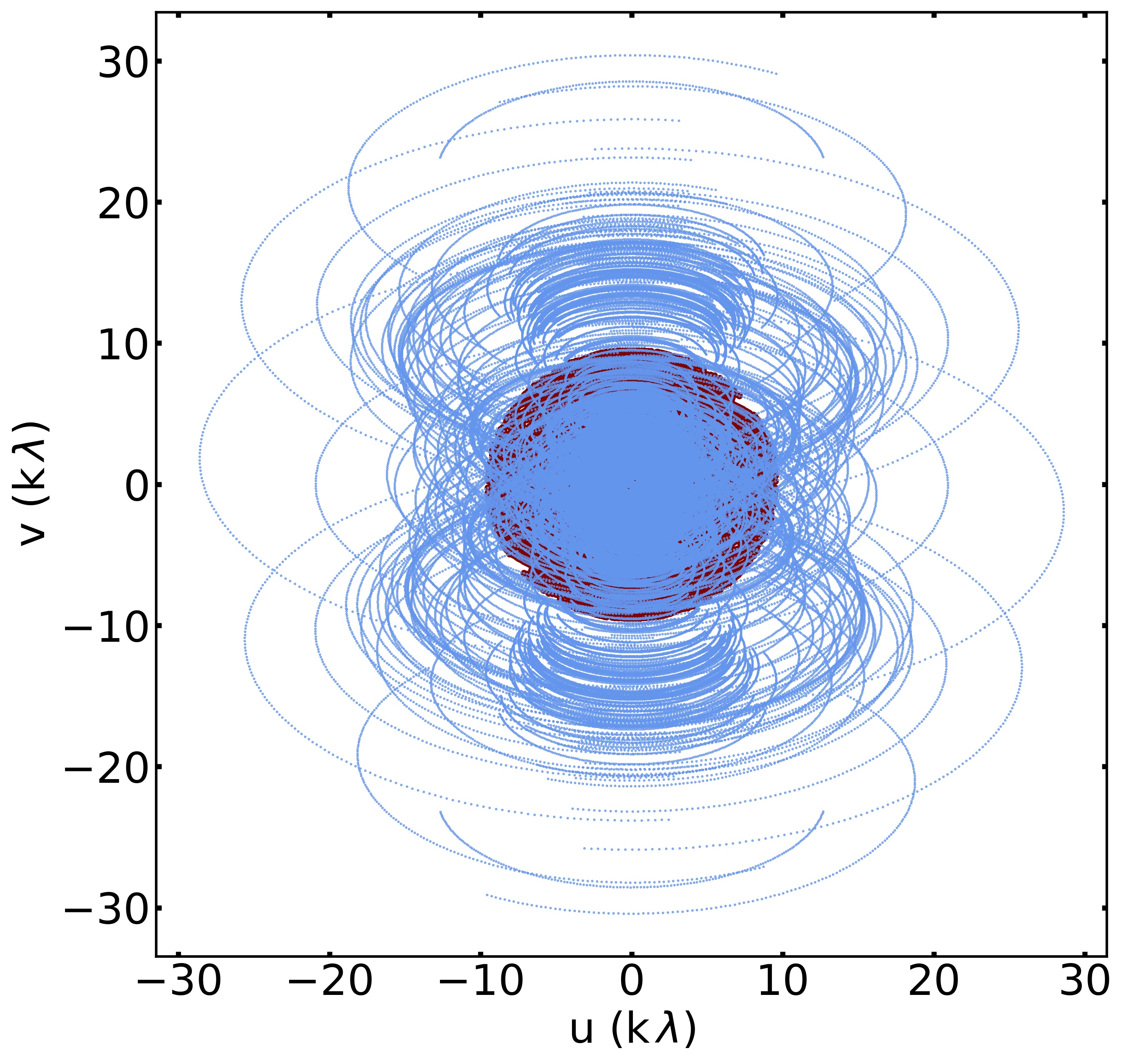}
			\caption{The uv-coverage for MeerKAT and ASKAP: The left panel represents the uv-coverage (all possible baselines) of the full MeerKAT array from source rise to source set for a source at declination $-30^{\circ}$ while retaining data for those times when that the source is at least $20^{\circ}$ above the horizon. The right panel represents the uv-coverage (sky blue) of the ASKAP array. The maroon-colored region shows a compact core within $2$ km, for the WALLABY survey at $-30^{\circ}$ declination.}\label{fig1}
		\end{figure*}
		
		\item
		The sensitivity of the telescope/survey is taken into account using
		the UV coverage for the array and other parameters of the
		telescope.  This, of course, is declination-dependent, in general.
		We choose to work with a declination that gives optimal
		sensitivity.
		Further, in order to account for galaxies that are resolved by the
		array, we use the sensitivity for an object of the angular size
		corresponding to the diameter of the galaxy: contribution from
		larger baselines is not taken into account. This is done as follows:
		
		We estimate the effective number of baselines (available antenna pairs) within a given distance using the Earth rotation synthesis. The UV-plane coverage for MeerKAT and ASKAP is shown in the figure \ref{fig1}. A radio interferometer measures the Fourier transform of the brightness distribution of the astrophysical source in the plane perpendicular to the direction of observation. For a given baseline, a point $(u,v)$ in the UV plane at a distance from the origin is equal to the projected length \citep{33} of the baseline measured in wavelength of observation. A baseline length $B$ is equal to $\sqrt{u^2+v^2}$. To estimate the effective number of baselines to compute the effective {\it rms} noise for a galaxy, we use a baseline length $B$ to estimate the angular resolution $\Delta\theta$ of the radio array and use the fact that the angular size $\theta_{size}$ of the galaxy is less than or equal to this angular resolution. The critical baseline $B_{crit}$ at which the size of the galaxy matches the angular resolution of the telescope.
		\begin{equation}
			\theta_{size} = 1.22\frac{\lambda}{B_{crit}}
		\end{equation}
		where $\theta_{size}= D_{HI}/D_{A}$ is the angular size of a galaxy with the galaxy size $D_{HI} $ and the angular diameter distance $D_{A}$.
		Those baselines that satisfy the $B \leq B_{crit}$ relation are considered for computing the sensitivity of the array for that galaxy. The number of these baselines is denoted by $N_B$. This relation simplifies the SNR calculation for galaxies that may otherwise be resolved by the array.  It is notable that such an approach will require making maps at different resolutions.  We also discuss an alternative approach below where we work with detection in maps constructed with different resolutions.
		
		\item
		In any given simulation, the location of each galaxy in the primary
		beam is used to compute the sensitivity as {\it rms} noise depends
		on the angle $\theta$ between the direction where the telescope is pointing
		and the object of interest. The primary beam pattern of an antenna can be modeled from a calibrator source, prior to observation. The attenuation pattern resulting from a cosine-tapered field \citep{34} of the MeerKAT antenna is used in our simulation. 
		\begin{equation}
			\beta(\theta) = \left[ \frac{cos(1.189\pi \theta/w)}{1-4(1.189\pi \theta/w)^2}\right]^{2}
		\end{equation}
		where $w$ is the full width half maximum (FWHM) of the antenna power pattern and it depends on the observed frequency $\nu$.
		\begin{equation}
			w = 57^{\prime}.5\left( \frac{\nu}{1.5\,\text{GHz}}\right)^{-1}
		\end{equation}
		ASKAP uses PAF technology (phased array feed) and forms $36$ beams simultaneously in the sky. 
		A Gaussian \citep{35} 
		shape is assumed to model the ASKAP primary beam. This model is parameterized by the width $w_{i}(\nu)$ and location $\theta_{i}$ of a particular beam from the boresight direction. This model suffers complications due to errors in beam position $\sigma_{\theta,i}$ and width $\sigma_{w,i}$. Taking these into account, the beam shape is as follows:
		\begin{equation}
			\beta_{i}(\theta) = \exp\left[- \frac{\left(\theta - \theta_{i}-\sigma_{\theta,i}    \right)^2}{2\left(1+\sigma_{w,i}\right)^2w_{i}(\nu)^2}\right]
		\end{equation}
		The authors \cite{36}
		showed from the holographic technique that the FWHM is $w(\nu)= 1.1c/\nu D$~rad where $D$ is the diameter of the antenna and $\nu$ is the observed frequency. The {\it rms} error in beam position $\sigma_{\theta,i}$ and error in width $\sigma_{w,i}$ are taken to be $1$ arcmin and $0.1$ arcmin respectively. The antenna's primary beam response reduces the signal from the source in the offset directions. In our simulations, we include the primary beam response of the antenna. The primary beam sensitivity is truncated at the $0.5$ sensitivity level in our simulations. Thus we use the primary beam upto half power.  This avoids the contribution of the antenna sidelobes. In the earlier prediction made by \citealt{20} for the MIGHTEE-HI survey and \citealt{22} for the WALLABY survey, the beam response function was not used, instead an average sensitivity was used for estimating the {\it rms} noise. 
		
		\item
		In general, the expected {\it rms} noise in the radio receiver system can be estimated using \citep{37}
		\begin{equation}
			\sigma_{rms} = \frac{T_{sys}}{G\sqrt{2N_{B}\,\Delta t\,\Delta \nu_{ch}}}
		\end{equation}
		where $T_{sys}$ is the system temperature, $\Delta t$ is the integration time,  $G=A_e/2k$ is the antenna gain, $A_{e}$ is the effective area of each antenna and $k=1380$ Jy.m$^2$ / K is the Boltzmann constant.
		
		\item
		The expected signal (flux density $S_{v}$) of a galaxy at redshift $z$, with HI mass $M_{HI}$ and velocity width $W_{20}$ can be computed using:
		\begin{equation}
			S_{v} \approx \frac{\beta(\theta)}{(2.36\times\,10^5\,W_{20})}\frac{M_{HI}\,(1+z)}{D^2_L}
		\end{equation}
		where $D_L$ is the luminosity distance of the galaxy and $\beta(\theta)$ is the primary beam pattern of the antenna. 
		The only approximation used here is that of a Gaussian line profile, which is different from a typical line profile on the redshifted $21$ cm line.
		
		\item
		Using all of these inputs, we were able to compute the SNR for each
		galaxy in the simulation that lies within the primary beam.
		We can then use this information to study the likelihood of blind/direct
		detection or contribution to detection with stacking.

		\item
		We have assumed the optimized signal-to-noise ratio by calculating the effective number of baselines when the galaxies are just unresolved. The derived image (when cleaning is applied on the real data) is a smoothed version of the true sky brightness due to the finite resolution of the array. The resolution of the image can be degraded depending upon the science we are interested in. This is useful for optimizing the signal-to-noise ratio for a given size of the sources on the map. The number of sources detected would also change when the resolution of the image is modified. We are interested in detected number counts on a particular angular resolution. We compute number counts of the detected galaxies on different angular resolutions, for instance, the native resolution of the array or twice that, by selecting the appropriate range of baselines. The native angular resolution is taken to be one pixel. The maximum angular resolution ( i.e. one resolution) used in this work is $\approx 8''$ for MeerKAT and it is $\approx 30''$ for ASKAP.
		
		We use projected area of the galaxy and the solid angle of the synthesized beam (converted into area) to compute the number of pixels across which a galaxy is spanned. The number of pixels $N_{pix}$ occupied by the galaxy for a given resolution (i.e. $1$ or $2$ resolution) can be computed as $N_{pix}= A_{g}/A_{Beam}$
		where, $A_g$ is projected area of the galaxy and $A_{Beam}$ is the area of the synthesized beam at a given resolution.
		The effective signal in Jy turns out to be $S_v/N_{pix}$. We estimate the number of detected galaxy in the map with a particular resolution using the SNR reduced by a factor of $1/N_{pix}$. The number counts for given resolution depends upon flux per pixel, rather than the full flux of the resolved galaxy.  We do this analysis for different effective resolutions while taking care to avoid the recounting detection of a given galaxy.

		Our predictions for MIGHTEE-HI based on the above simulations have been validated with previous work.  This has been done by adopting the method used by \citealt{20} to estimate various quantities. They took an {\it rms} noise $\approx 100\mu$Jy/beam per channel and a linewidth $\approx 150$ km/s (independent of the HI mass) and predicted detection of around $3000$ galaxies with $SNR \geq 5$ in the full MIGHTEE-HI survey with a typical field of view. 
		
		Our predictions for WALLABY number counts have also been validated by comparing with the earlier work by \citealt{22} with our model overestimating the total number counts by $\approx 2\%$. 
		To validate these results with the earlier work, we took the following scaling relations: mass-circular velocity, dispersion in linewidth, mass-diameter relation, and frequency resolution, as used therein. These validations are presented in table \ref{Tab2}. We do not take clustering into account when predicting number counts.
		Though it is shown in \citealt{26} that clustering affects the HI mass function. They presented the HIMF calculation of the local universe for the spring sky and the fall sky. The spring sky is dominated by the supercluster, a very dense region of objects. The fall sky region is facing towards the local void. They found that the low-mass end slope is much steeper in the spring sky than the fall sky. 
		However, the volume being probed in the proposed survey is much larger and the impact of inhomogeneities in the galaxy distribution is likely to be much smaller. 
		However, the predictions of the large-scale surveys can get offset due to the impact of local inhomogeneities as we anchor our predictions in the local measurement of the HIMF. 
		We study the impact of varying the HIMF parameters in \S{4}. 
	\end{itemize}
	\begin{table}[ht]
		
		\tabularfont
		\setlength{\tabcolsep}{5pt}
		
		\caption{A comparison between earlier predictions and predictions made by the method used in this work: The number of galaxies detected with $5$-$\sigma$ confidence in the HI surveys is presented in this table.}\label{Tab2} 
		\begin{tabular}{lccccc}
			\topline
			HI-Surveys& Earlier prediction & Our predictions\\
			\midline
			MIGHTEE-HI&         $3000$   &       $3110\pm 57$      \\
			WALLABY&            $7.3\times 10^5$&      $7.48\pm0.06\times 10^5$      \\
			
			\hline
		\end{tabular}
	\end{table}

	\section{Parameters of MIGHTEE-HI and WALLABY}
	
	The MIGHTEE-HI survey is an international collaboration project with the MeerKAT\footnote{\url{https://www.sarao.ac.za/gallery/meerkat/}} array. The smaller aperture of the MeerKAT offers a larger FoV than uGMRT and VLA.  MeerKAT has a single-pixel feed with $64$ Gregorian dishes of size $13.5$~m that provide a $\,\approx1$ deg$^2$ field of view at $1420$ MHz ($z=0$). The maximum baseline length is $7.5$ km, corresponding to the resolution $\approx8$ arcsec at $z=0$. It will observe an area of approximately 32 deg$^2 $ of the southern sky that includes the Fornax cluster as part of the MIGHTEE-HI survey. We take $30$ pointings (i.e. 30 fields) in our simulations, which are equivalent to 32 deg$^2 $ area allocated to the MIGHTEE-HI survey \citep{38}. The number of pointings may change in the actual observations, depending on the mosaic pattern for a particular field.  In this work, we assume that the pointings are independent.  Although our work is focused on the MIGHTEE-HI and WALLABY surveys and we do not make predictions for LADUMA in this work. LADUMA is a very deep survey and will be carried out by MeerKAT with $1000$ hrs per pointing \citep{39}. LADUMA is the deepest HI-survey planned with SKA precursors, and it will observe a single field of Extended Chandra Deep Field South (ECDF-S) for approx. $4000$~hrs. MIGHTEE-HI will detect more high-HI-mass galaxies at low redshift, and LADUMA with $~4.5$ times higher sensitivity will detect more low-HI-mass galaxies. The combination of these two will enable us to constrain both the low-mass and high-mass ends of the HI mass function.
	
	\begin{table}[ht]
		
		\tabularfont
		\setlength{\tabcolsep}{2pt}
		
		\caption{The key parameters of the surveys}\label{Tab3} 
		\begin{tabular}{lccccc}
			\topline
			Parameters& MIGHTEE-HI &WALLABY&   Unit\\
			\midline
			maximum baseline&            7.5&           2&     km\\
			angular resolution&          8-80&         30&     arcsec\\
			survey area&            32&           30940&     deg$^2$\\
			number of pointings&            30&           1200\textsuperscript{*}&     \\
			integration time/pointing&            25&         8&     hours\\
			system temperature&            20&        50&     K\\
			redshift range&            0-0.4&           0-0.26&     \\
			bandwidth&            850&          300&     MHz\\
			velocity resolution&            5.5&          4&     km/s\\
			rms sensitivity&            100&          700&     $\mu$Jy\\
			\hline
		\end{tabular}
		\tablenotes{Some ASKAP pointings may overlap to cover allocated area\textsuperscript{*}}
	\end{table}

	ASKAP\footnote{\url{https://www.csiro.au/en/about/facilities-collections/ATNF/ASKAP-radio-telescope}} is a SKA pathfinder telescope and has $36$ dishes, each with a aperture $12$ m. The maximum baseline length is $6$ km. The inner core consists of $30$ antenna within $2$ km area that provides a dense core designed for high sensitivity. The ASKAP will observe $1200$ pointing during the $9600$ hrs allocated for WALLABY. It is expected to detect approximately $10^5$ galaxies at low redshift. This will help to constrain HIMF and other aspects of cosmology \citep{40} (e.g. Dark Energy equation of state). WALLABY will probe $3/4$ of the sky ($\delta < +30^{\circ} $) to the redshift of $z\leq 0.26$. WALLABY survey has a flux sensitivity $20$ times better than the HIPASS survey and will detect galaxies with $2$ order of magnitude higher than the HIPASS survey. The other key parameters of the survey are given in the table \ref{Tab3}.

	\section{Predictions}
	
\begin{figure*}
		\centering\includegraphics[width=.3\textheight]{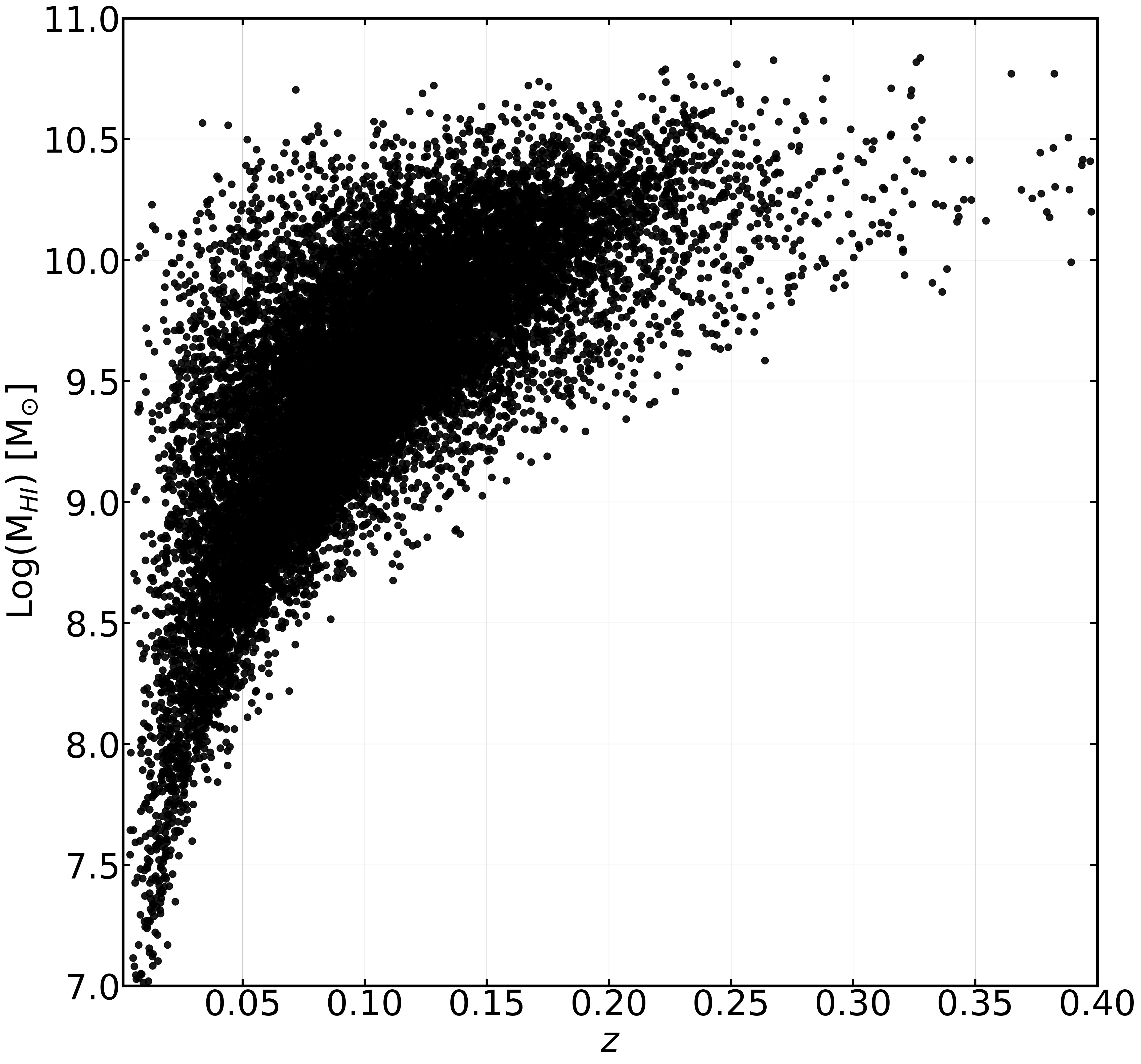}
		\centering\includegraphics[width=.3\textheight]{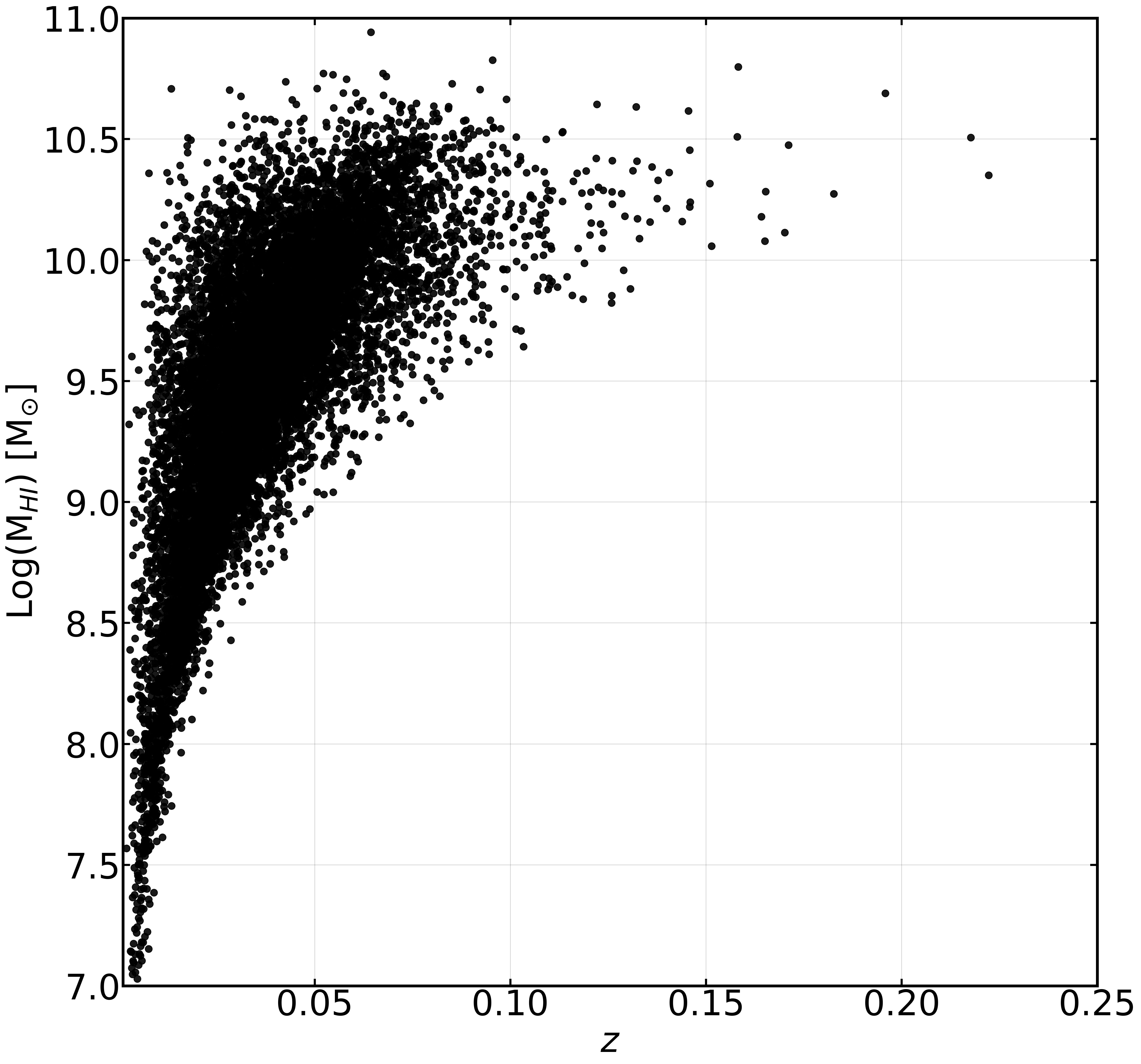}
		\caption{Mass redshift scatter plot from one simulation for MIGHTEE-HI and WALLABY: The left panel represents the distribution of galaxies detected with SNR$>5$ for MIGHTEE-HI. These are the total galaxies in Model III detected within the $30$ pointings. The right panel shows the distribution of galaxies in mass redshift space detected within the $15$ pointings (i.e. smaller sky area than full survey) of the WALLABY survey with SNR$>5$. The knee mass $M^*=10^{9.96}$ galaxies are observed out  to redshift $z \leq 0.22$ and $z \leq 0.09$ by the MIGHTEE-HI and WALLABY surveys, respectively.}\label{fig2}
	\end{figure*}

	In this section, we present the results based on the methodology discussed in section 2. We assume the ALFALFA HIMF parameters for our fiducial case: low-mass end power law index $\alpha=-1.33$, knee mass $M^*_{HI}=10^{9.96}$ $M_{\odot}$ and normalization $\phi^*= 4.18\times 10^{-03}$ Mpc$^{-3}$. Parameter values (ALFALFA HIMF \citealt{1}) are varied by $\pm 5\%$  and $\pm 10\%$  to study variations in the predicted number of detections. Instead of directly changing $\phi^*$, we change $\Omega_{HI}$, and use $\phi^*$ to normalize the HIMF. We used logarithmic relations between variation of number counts and specific HIMF parameters to check the sensitivity of parameters against the number of detected galaxies. The sensitivity for a HI mass function to the MIGHTEE-HI survey was explored earlier in Fig. 3 of \cite{20}. This was done by counting the number of galaxies in the cells of mass-redshift space, similar to the figure \ref{fig2} in this work. Their sensitivity investigation was towards one HIMF without variation in parameters. In our work, we also explore the sensitivity of number counts with respect to multiple HIMFs by varying the parameters.
	It is important to note that the HIMF will be measured from the observations. This measurement will make use of complete details available in the observations. In this paper, we are exploring the variation in number counts with the HIMF as a gross indicator.
	We computed mock catalogs for the three models, as discussed earlier. The left panel of figure \ref{fig2} shows HI mass-redshift space distribution of galaxies within $30$ independent pointings of the MIGHTEE-HI survey, which are detected with an SNR greater than $5$ in $25$ hours of integration time. A larger number of galaxies will be detected at low redshifts with MIGHTEE-HI, but the regions ($z: 0.09-0.21 $) are affected by RFI (N. Madox {\em et al.}2020). This can have an impact on the total number of galaxies detected in the actual survey, but we do not take this into account here. Model III is taken to produce the mass-redshift space in the figure \ref{fig2}.

	\begin{figure*}
		\centering\includegraphics[width=.3\textheight]{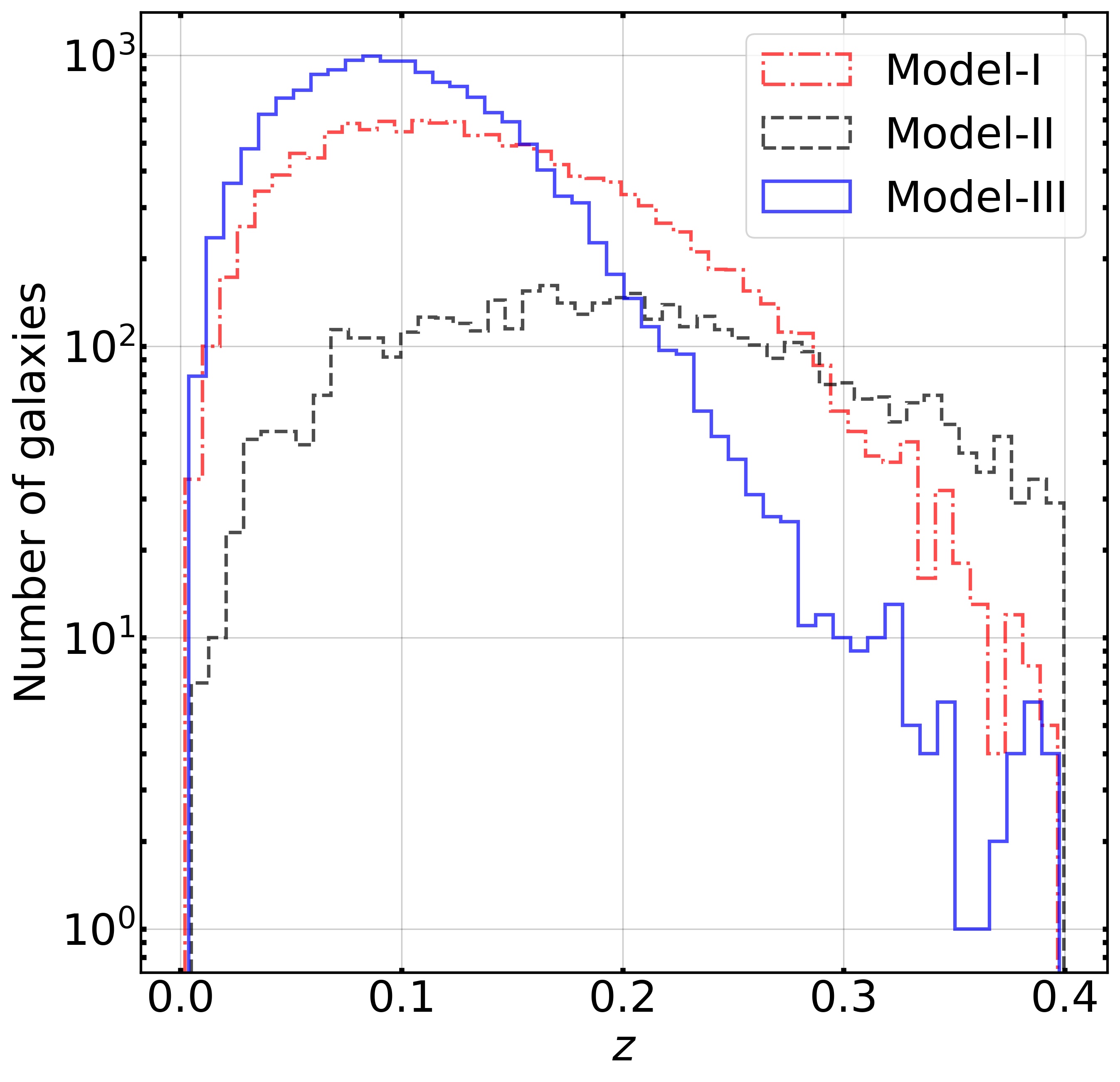}
		\centering\includegraphics[width=.3\textheight]{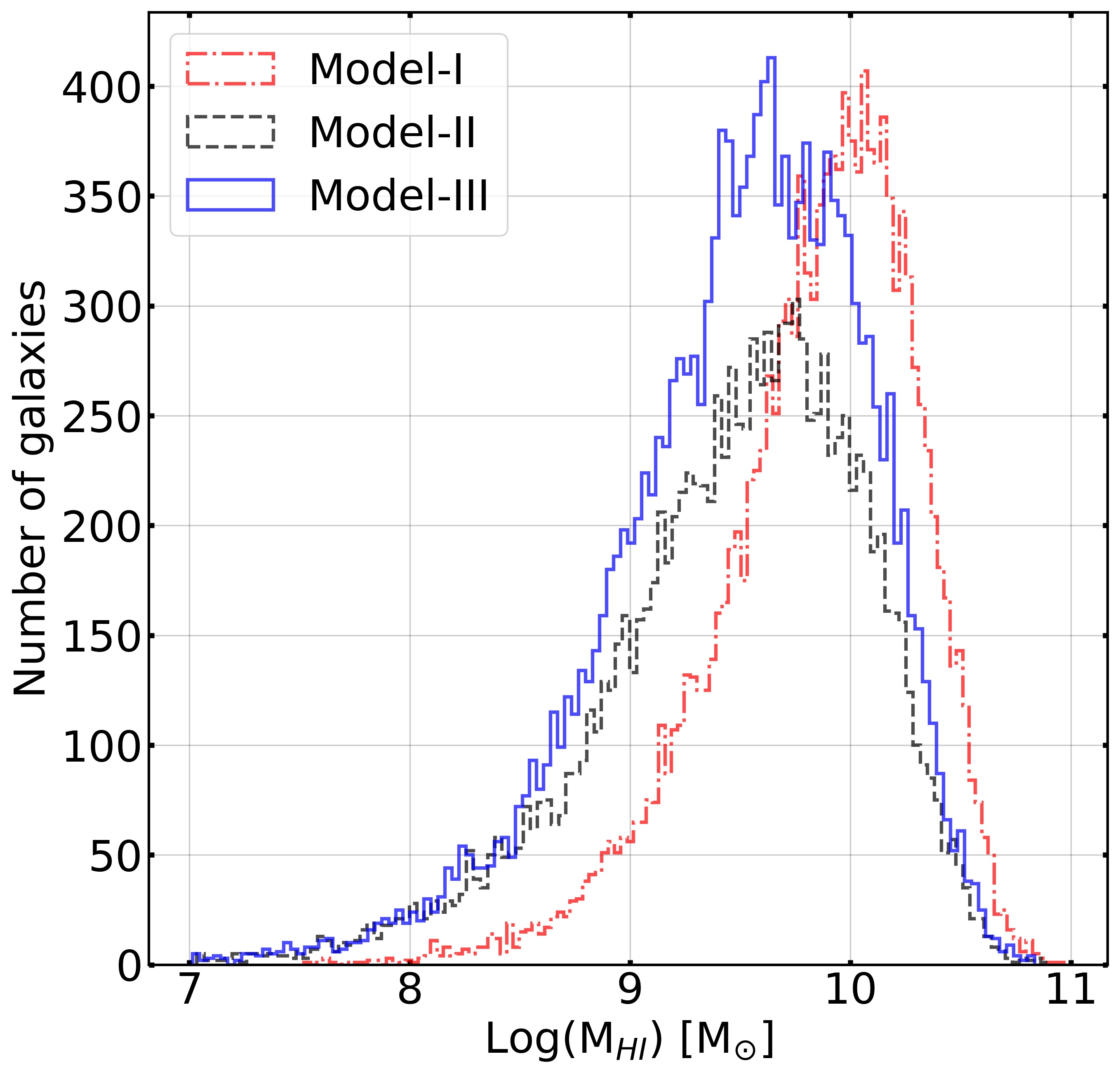}
		\caption{Distribution of galaxies across the HI mass and redshift: In the left panel, blind detection of galaxies as a function of redshifts is shown for three models ( for the full MIGHTEE-HI survey). The right panel represents the HI mass distribution of galaxies from our mock catalogs with SNR$> 5$. In Model I, we take a fixed size and a fixed rotation velocity. Model II assumes mass-dependent sizes and rotation velocities of galaxies. Model III accounts for mass-dependent HI diameter and rotation velocity with random orientations of galaxies.}\label{fig3}
	\end{figure*}
	
	\begin{figure*}[!hbt]
		\centering\includegraphics[width=.3\textheight]{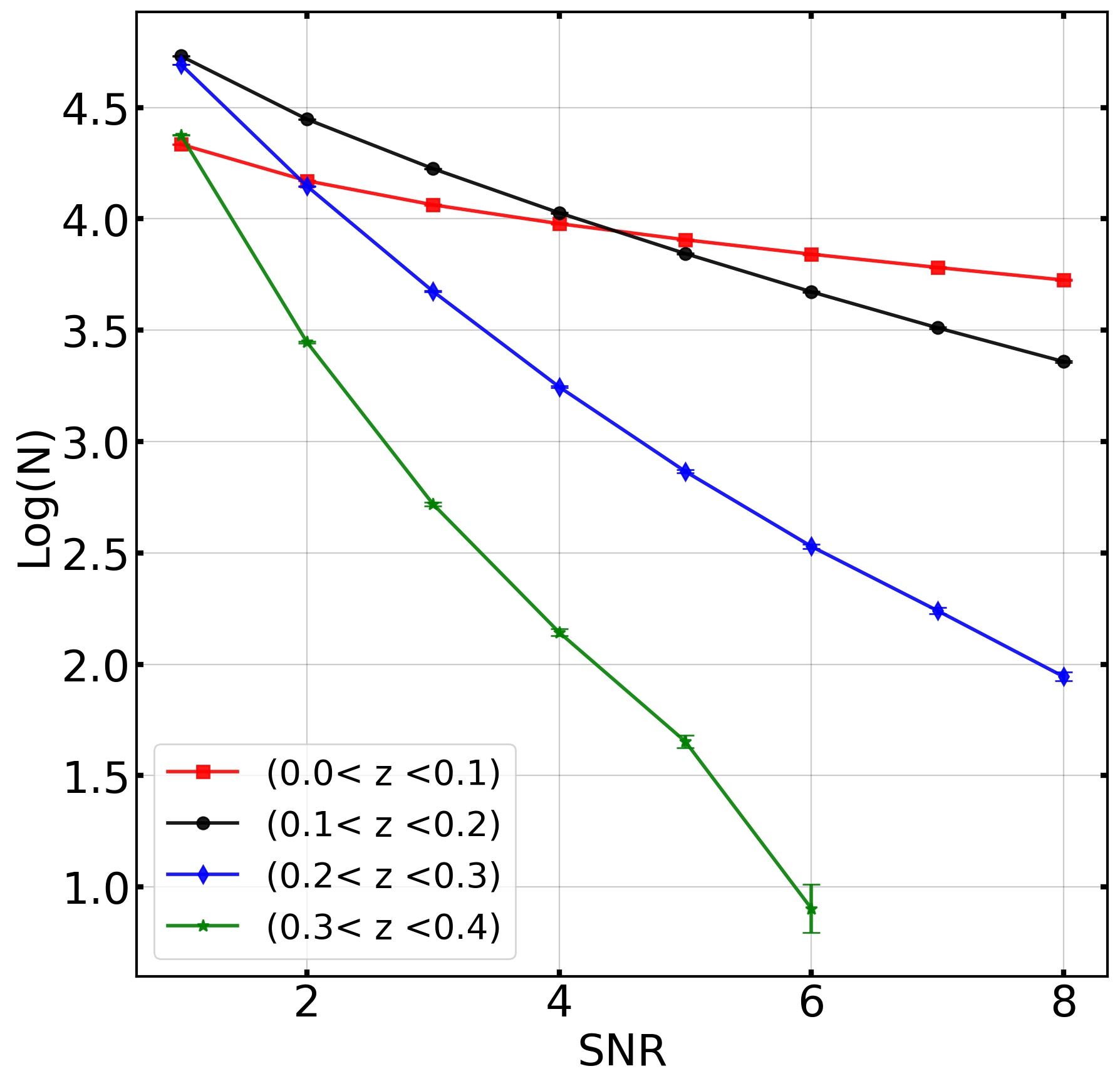}
		\centering\includegraphics[width=.3\textheight]{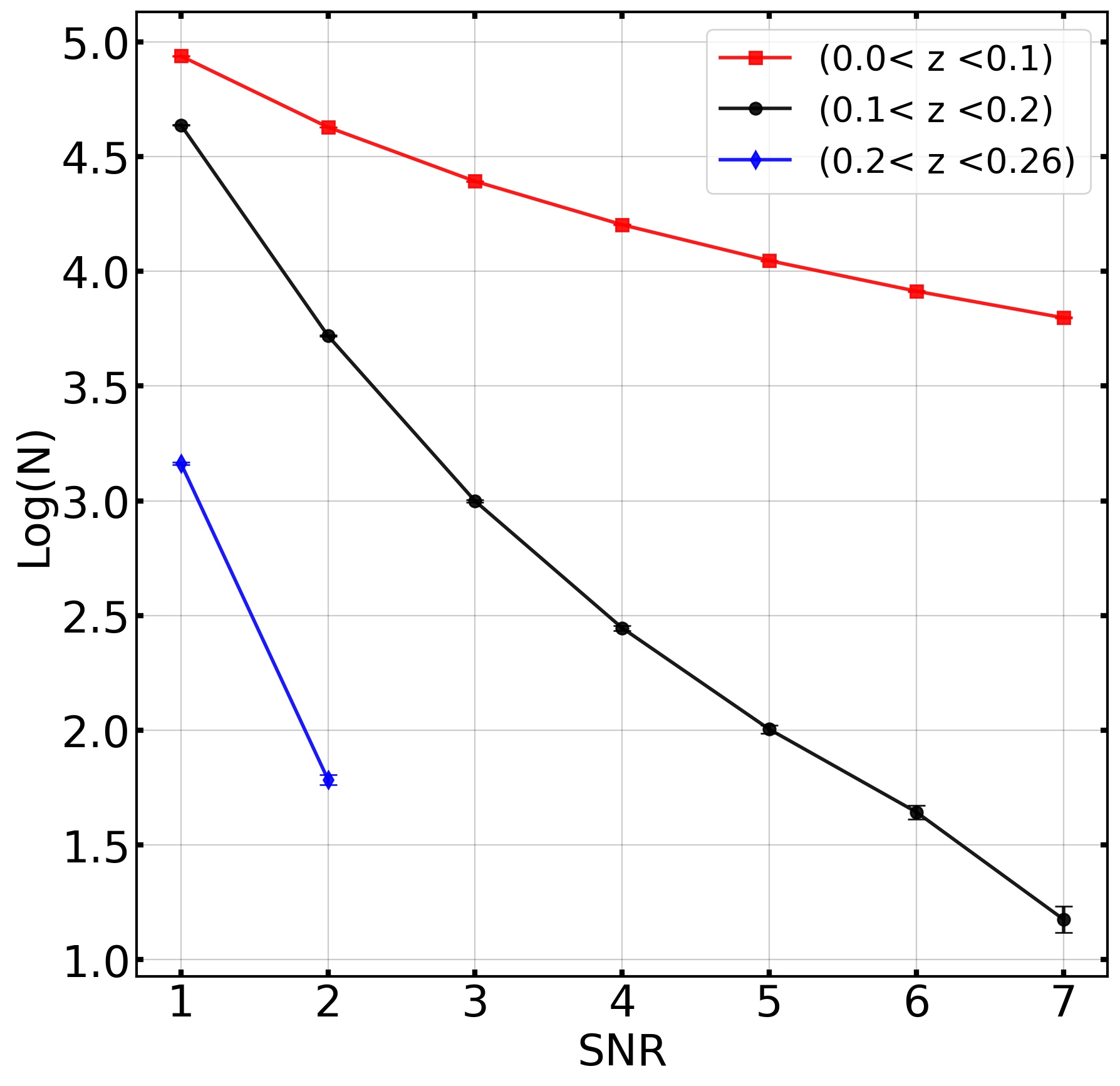}
		
		\caption{Number counts and signal-to-noise ratio (SNR): The left panel represents the number counts as a function of signal-to-noise ratio in the full MIGHTEE-HI survey (30 pointings) for Model III. These numbers take the dispersion across simulations into account with $3\sigma$ error bars.  The ensemble of mock catalogs used $50$ realizations per pointing.  Approx. $45$ galaxies are detected with SNR $> 5$ in the $0.3>z>0.4$ bin. The right panel represents the same for $15$ ASKAP pointings in the WALLABY survey. The full WALLABY survey area is $30940$ deg$^2$ that can be probed with $\approx1000$ ASKAP's pointings. A single pointing area is $\approx 30$ deg$^2$. To obtain the total number counts in the right panel, one needs to scale these number counts with a scale factor of $30940/(15\times30)\approx 68$.}\label{fig4}
	\end{figure*}
	
	\begin{figure*}
		
		\centering\includegraphics[width=.3\textheight]{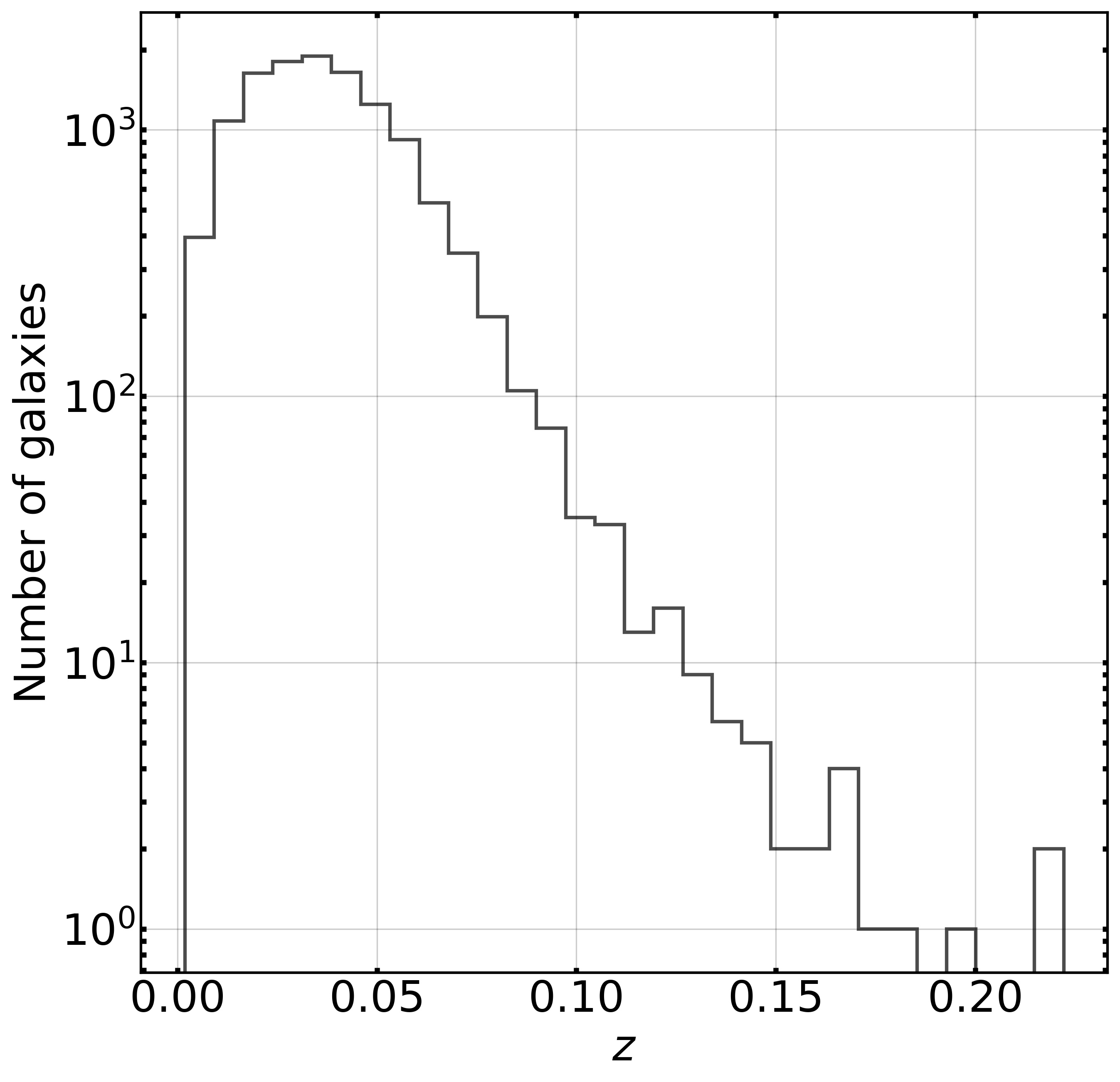}
		\centering\includegraphics[width=.3\textheight]{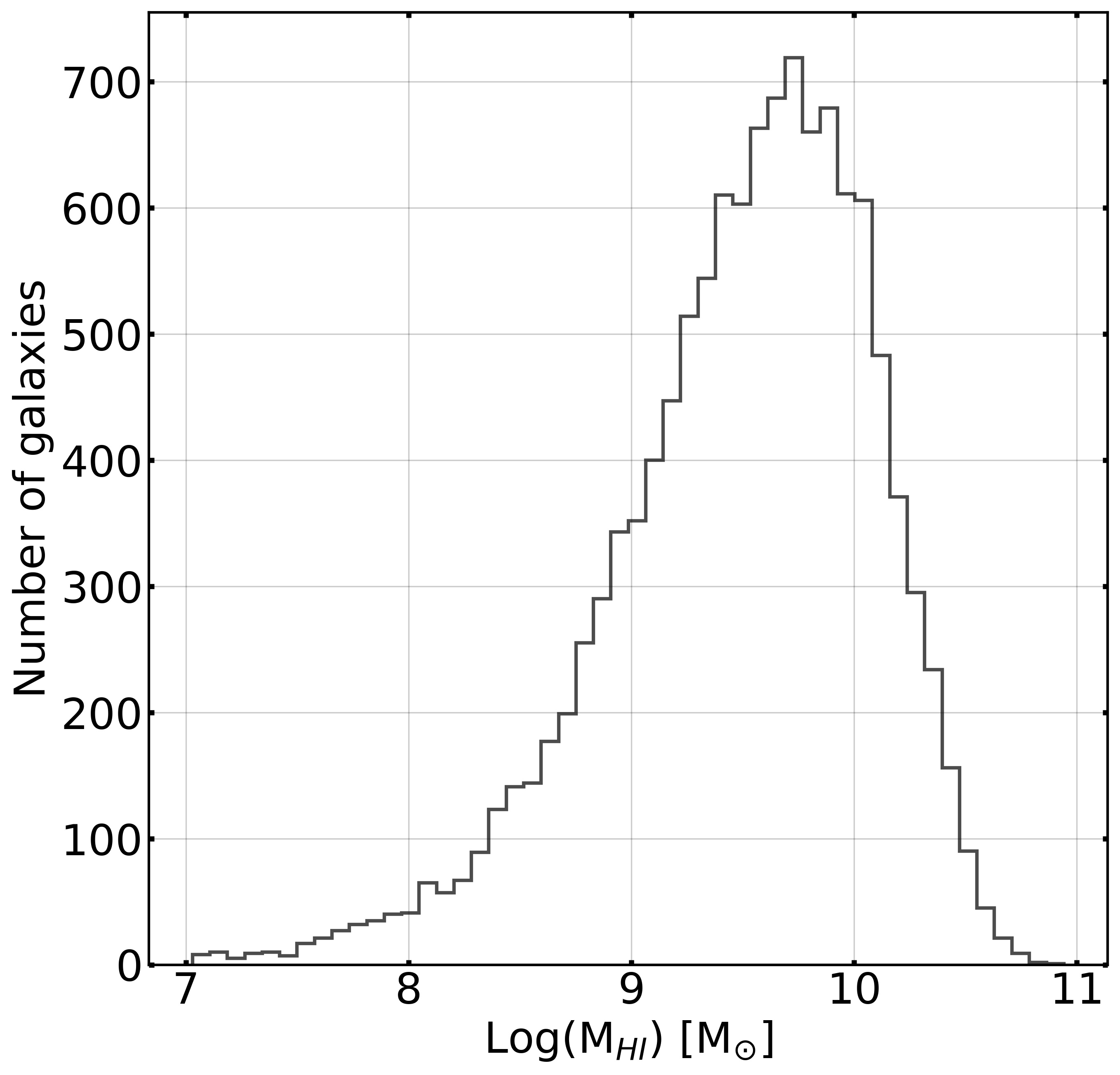}
		\caption{
			Distribution of galaxies for the WALLABY survey (15 pointings, i.e sky area smaller than full WALLABY ): In the left panel, a blind detection of galaxies as a function of redshifts is shown. The right panel represents the HI mass distribution of detected galaxies. Both the distributions are at above $5$-$\sigma$ level. In order to get full WALLABY number counts, One needs to multiply these number with scale factor of $68$.}\label{fig5}
	\end{figure*}

	The distribution of galaxies in the mock catalog with SNR $\geq5$, as a function of redshift in the full MIGHTEE-HI survey, is shown in the left panel of the figure \ref{fig3}. These are the total number of galaxies detected up to $z\approx 0.4$. In the right panel of figure \ref{fig3}, the expected distribution of the detected galaxies as a function of the HI mass is shown. 
	The overall trend has some common features in all three models: There is a rise at low redshifts as the volume in the observed solid angle increases with the redshift, followed by a decline as galaxies with low HI mass can no longer be detected. 
	The highest number of predicted detections is for Model-III: especially the counts at low redshifts are the highest here.  
	This is a combination of mass-dependent size, linewidth, and random orientation.  
	The mass-dependent size leads to a smaller size for low-mass galaxies, and the number of baselines that can be used increases.
	At the same time, random orientation leads to a smaller linewidth and a larger peak flux that makes detection more probable even than that of Model II (inclination ignored, i.e. edge-on). In the figure \ref{fig3}, Model-I with fixed size ($50$ kpc) and velocity width ($300$ km/s) has a smooth trend for redshift distribution (initial rise followed by decline due to limitations of sensitivity at higher redshifts). In the HI mass-dependent Model-II, galaxy size and linewidth decrease for low-mass galaxies (unlike Model I that catalogs all galaxies with same size and linewidth) at low redshifts, and resulting lower count in comparison to Model I. Model III uses the inclination that makes the linewidth even much smaller, therefore signal get peaked at low redshifts that leads to highest number counts.

	\begin{table*}
		\setlength{\tabcolsep}{3pt}
		\centering
		\tabularfont
		\caption{The sensitivity of the HI mass function: In this table, the slopes of the logarithmic relation between the number of detections and parameters (Figures \ref{fig6} and \ref{fig7}) are presented for the MIGHTEE-HI survey. 
			The first column is a description of the quantity in that row with a parathentical description of any cuts that may have been applied.
			In columns 2-5, slopes and uncertainties for the MIGHTEE-HI survey are presented for each redshift bin. The numbers in brackets represent the galaxies at ALFALFA values of the HI mass function parameter, corresponding to different cuts and redshift bin. Some entries are empty because there have been no predicted detections of galaxies in the corresponding redshift bin. 
			It is noteworthy that the slope of the $\log(N)-\log(\Omega_{HI})$ relation does not vary significantly as a function of redshift (except  $0.3<z<0.4$) and different cuts; therefore, the HI mass density parameter is not sensitive to the number of galaxies detected at lower redshifts. The slope of variation with $\alpha$ changes sign from redshift window $0<z<0.1$ to $0.1<z<0.2$: this is a signature of variation in slope.  The slopes have been derived using least square fit.}\label{Tab4}
		\begin{tabular}{lccccccr}
			\topline
			Relation & $0<z<0.1$ &$0.1<z<0.2$&$0.2<z<0.3$&$0.3<z<0.4$\\
			\midline
			
			$\log(N)-\log(\alpha) $&$0.90\pm0.03 (8053)$ & $-1.18\pm0.24 (6980)$ & $-2.32\pm0.31 (734)$&$-3.97\pm2.37 (45)$\\
			
			$\log(N)-\log(\alpha)$ ($W_{20} > 200$ km/s)&$-0.77\pm0.26(2539)$&$-1.4\pm0.27(4906)$&$-2.87\pm0.40(414)$& \\
			$\log(N)-\log(\alpha)$ ($W_{20} < 200$ km/s)&$1.68\pm0.04(5505)$&$-0.35\pm0.2(2058)$&$-1.72\pm0.33(305)$& $-2.45\pm0.76(31)$\\
			
			$\log(N)-\log(\alpha)$ ($M_{HI} > 10^{9.5}M_{\odot}$)&$-1.10\pm0.24(2052)$&$-1.49\pm0.27(5765)$&$-2.30\pm0.29(727) $&$-3.97\pm2.3(45)$\\
			
			$\log(N)-\log(\alpha)$ ($M_{HI} < 10^{9.5}M_{\odot}$)&$1.59\pm0.07(5988)$&$0.27\pm0.2(1198)$& &\\
			
			$\log(N)-\log(M^*) $ &$-0.36\pm0.01$ & $0.24\pm0.03$ & $1.30\pm0.15$& $3.22\pm0.86$\\
			
			$\log(N)-\log(M^*)$  ($W_{20} > 200$ km/s)&$10^{-3}\pm0.05$&$0.43\pm0.05$&$1.96\pm0.2$& \\
			
			$\log(N)-\log(M^*)$  ($W_{20} <200$ km/s)&$-0.52\pm0.003$&$-0.21\pm0.6$&$0.48\pm0.22$&$0.70\pm0.50$\\
			$\log(N)-\log(M^*)$  ($M_{HI} > 10^{9.5} M_{\odot}$)&$0.12\pm0.09$&$0.38\pm0.05$&$1.30\pm0.16$&$3.22\pm0.86 $\\
			$\log(N)-\log(M^*)$ ($M_{HI} < 10^{9.5}M_{\odot}$)&$-0.52\pm0.01$&$-0.45\pm0.04$& & \\
			$\log(N)-\log(\Omega_{HI})$&$1.01\pm0.01$&$0.99\pm0.04$&$1.03\pm0.18$&$2.43\pm1.21$\\
			$\log(N)-\log(\Omega_{HI})$ ($W_{20} > 200$ km/s)&$1.02\pm0.03$&$0.98\pm0.05$&$1.04\pm0.12$&  \\
			$\log(N)-\log(\Omega_{HI})$ ($W_{20} < 200$ km/s)&$1.01\pm0.03$&$1.01\pm0.05$&$1.04\pm0.20$& $0.66\pm0.
			86$ \\
			$\log(N)-\log(\Omega_{HI})$ $(M_{HI} > 10^{9.5}M_{\odot})$ &$0.99\pm0.05$&$1.0\pm0.04$&$1.02\pm0.17$&$2.43\pm1.21$\\
			$\log(N)-\log(\Omega_{HI})$ ($M_{HI} < 10^{9.5}M_{\odot}$)&$1.02\pm0.02$&$0.97\pm0.05$&  & \\
			\hline
		\end{tabular}
		
	\end{table*}
	
	\begin{figure*}[!hbt]
		\centering\includegraphics[width=.32\textheight]{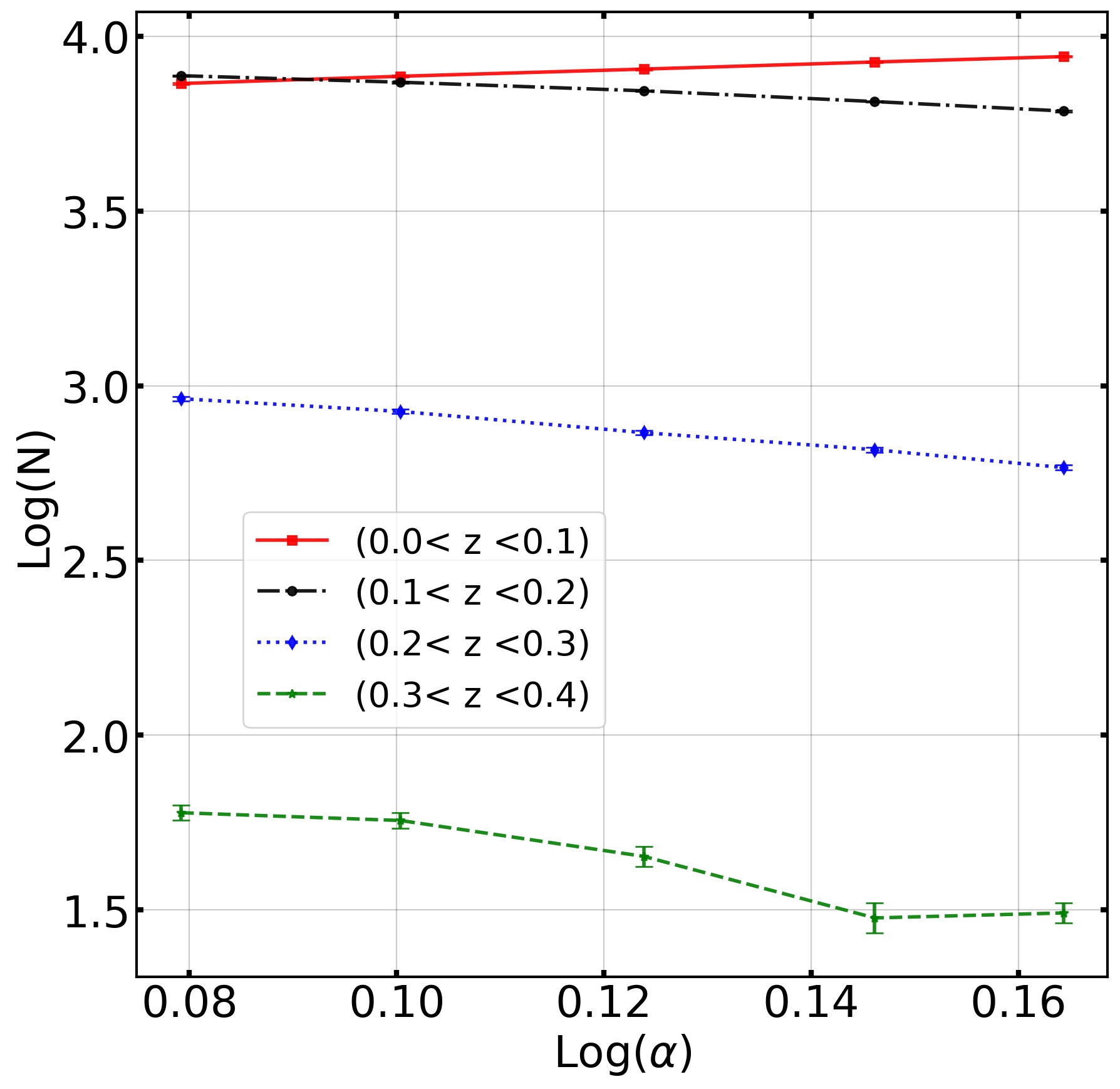}
		\centering\includegraphics[width=.3145\textheight]{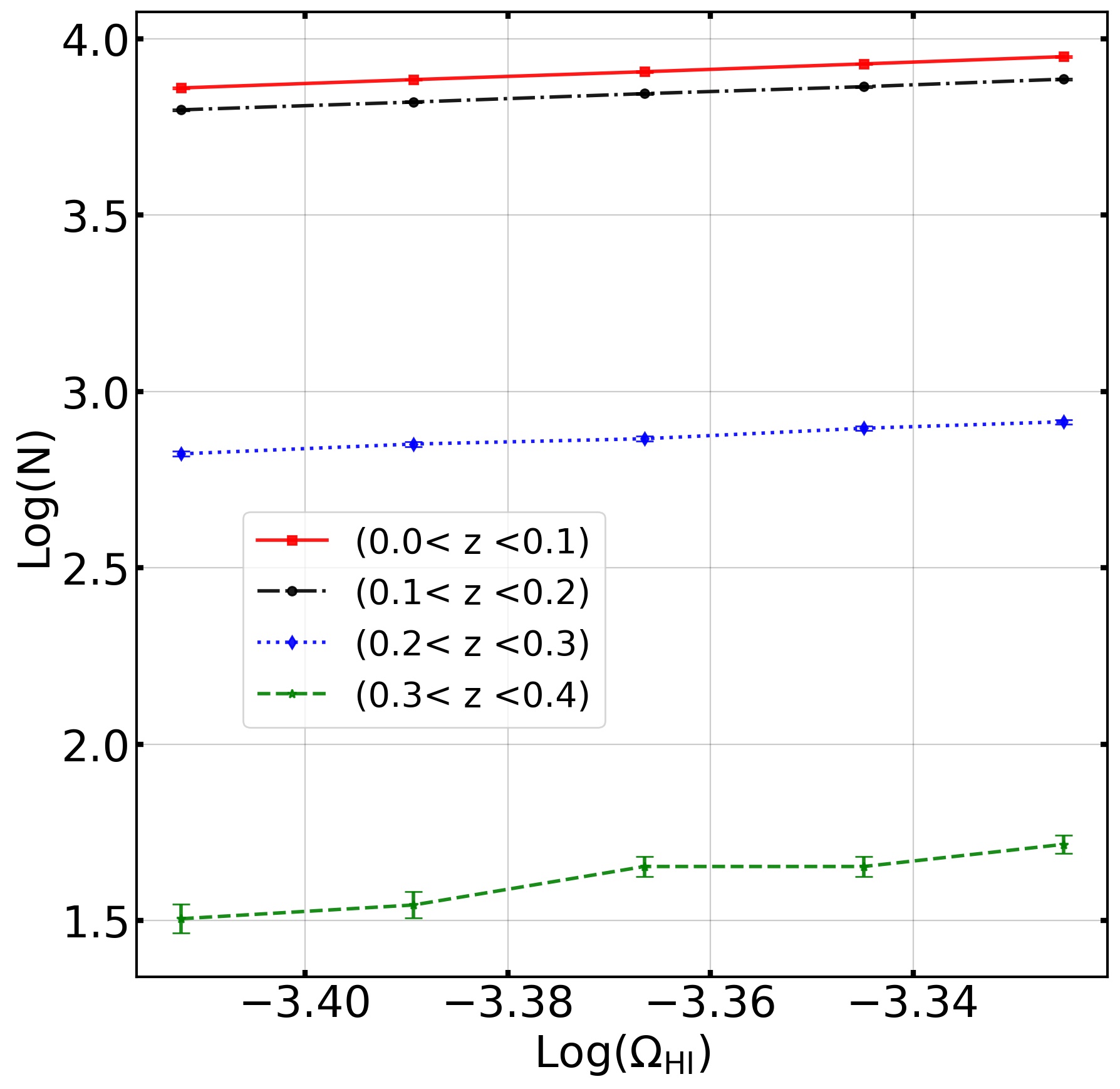}
		\caption{Sensitivity to HIMF parameters for the MIGHTEE-HI survey: The left panel represents the number of galaxies in the mock catalogs with SNR $> 5$ as a function of the low mass end slope $\alpha$ for each of the four redshift bins. The x-axis of both panels indicates $5$ values of the HIMF parameter. The middle value on the x-axis represents the ALFALFA value and others points are $\pm5\%$ and $\pm10\%$ from the ALFALFA value. The slope inversion for the lowest redshift bin as compared to other bins can be seen here. The right panel represents the same but variation of numbers with $\Omega_{HI}$ HI mass density parameter.}\label{fig6}
	\end{figure*}

	The cumulative number of estimated detections as a function of SNR is shown in figure \ref{fig4} for MIGHTEE-HI and WALLABY surveys. 
	We takel $30$ independent MeerKAT pointings (MIGHTEE-HI) each with area $\approx 1$ deg$^2$. 
	These are the expected numbers of detection with $3\sigma$ error bars of 50 realizations per pointing. The total predicted numbers ($30$ pointings) are computed assuming no overlap between fields.  Please note that some of the fields allocated for the MIGHTEE-HI survey will be observed with an overlap strategy \citep{41}. In case of overlap in fields, the total sky coverage decreases but the sensitivity in overlap regions is higher.  Details of the observation strategy can be used to make refined predictions. 
	
	The MIGHTEE-HI survey, being deeper, is more sensitive to low-HI mass galaxies. The overall distribution of detections follows a pattern dictated by increasing volume with redshift, followed by sensitivity limitation that leads to a drop in the number of detections at higher redshifts.  
	This can be seen clearly in the left panels of figures \ref{fig3} and \ref{fig4}.
	This also impacts the distribution of HI masses. 
	As we can detect low-HI mass galaxies only at low redshifts where the survey volume is small, the number of such galaxies that can be detected is small.
	As we move to higher HI masses, the redshift up to which these can be detected increases, and the numbers increase. 
	This trend continues till we get to masses that are comparable with the knee mass $M^*$ as the number density of galaxies with a higher HI mass drops sharply and hence the total number of detections also drops sharply. In figure \ref{fig2}, it is shown that galaxies around the knee mass can be observed up to redshifts $z=0.22$ and $z=0.09$ for the MIGHTEE-HI and WALLABY surveys, respectively. We are able to detect only more massive galaxies at higher redshifts.  Therefore, completeness of the sample is an important issue for calculating the HIMF, e.g., with the 2DSWML method \citep{42}. 
	
	\begin{table*}
		\setlength{\tabcolsep}{3pt}
		\centering
		\tabularfont
		\caption{Sensitivity to the HIMF for the WALLABY survey: The slopes of the logarithmic relations between the number of detections and the parameters are presented in this table. The number in the bracket represents the galaxies at ALFALFA values of the HI mass function parameter that corresponds to different cuts and the redshift bin. The relations are given in the first column with and without cuts in the HI mass ($10^{9.5} M_{\odot}$) and linewidth $(200)$ km/s. The slopes are given in columns 2-3.}\label{Tab5}
		\begin{tabular}{lccccccr}
			\topline
			
			Relation &$0<z<0.1$ & $0.1<z<0.2$\\
			\midline
			$\log(N)-\log(\alpha)$ &$-0.18\pm0.08(11118)$&$-2.76\pm0.46(101)$\\
			$\log(N)-\log(\alpha)$ ($W_{20} > 200$ km/s)&$-1.3\pm0.25(5822)$&$-10.5\pm12.57(13)$\\
			$\log(N)-\log(\alpha)$ ($W_{20} < 200$ km/s)&$0.94\pm0.06(6026)$&$-2.5\pm0.25(82)$\\
			$\log(N)-\log(\alpha)$ ($M_{HI} > 10^{9.5}M_{\odot}$)&$-1.45\pm0.27(6431)$&$-2.76\pm0.46(101)$ \\
			$\log(N)-\log(\alpha)$($M_{HI} < 10^{9.5}M_{\odot}$) &$1.32\pm0.07(5418)$& \\
			$\log(N)-\log(M^*) $&$-0.2\pm0.02$& $1.45\pm0.43$\\
			$\log(N)-\log(M^*)$  ($W_{20} > 200$ km/s)&$0.36\pm0.04$&$8.71\pm7.12$\\
			$\log(N)-\log(M^*)$  ($W_{20} <200$ km/s)&$-0.41\pm0.01$&$1.13\pm0.32$\\
			$\log(N)-\log(M^*)$  ($M_{HI} > 10^{9.5} M_{\odot}$)&$0.38\pm0.02$&$1.45\pm0.43$\\
			$\log(N)-\log(M^*)$ ($M_{HI} < 10^{9.5}M_{\odot}$)&$-0.52\pm0.02$& \\
			$\log(N)-\log(\Omega_{HI})$&$1.0\pm0.03$&$1.0\pm0.24$\\
			$\log(N)-\log(\Omega_{HI})$ ($W_{20} > 200$ km/s)&$1.0\pm0.05$&$1.17\pm1.13$\\
			$\log(N)-\log(\Omega_{HI})$ ($W_{20} < 200$ km/s)&$1.0\pm0.03$&$1.0\pm0.3$\\
			$\log(N)-\log(\Omega_{HI})$ $(M_{HI} > 10^{9.5}M_{\odot})$&$1.0\pm0.05$&$1.03\pm0.24$ \\
			$\log(N)-\log(\Omega_{HI})$ ($M_{HI} < 10^{9.5}M_{\odot}$) &$1.0\pm0.02$&\\
			\hline
		\end{tabular}
	\end{table*}
	
	\begin{figure*}[!hbt]
		\centering\includegraphics[width=.85\columnwidth]{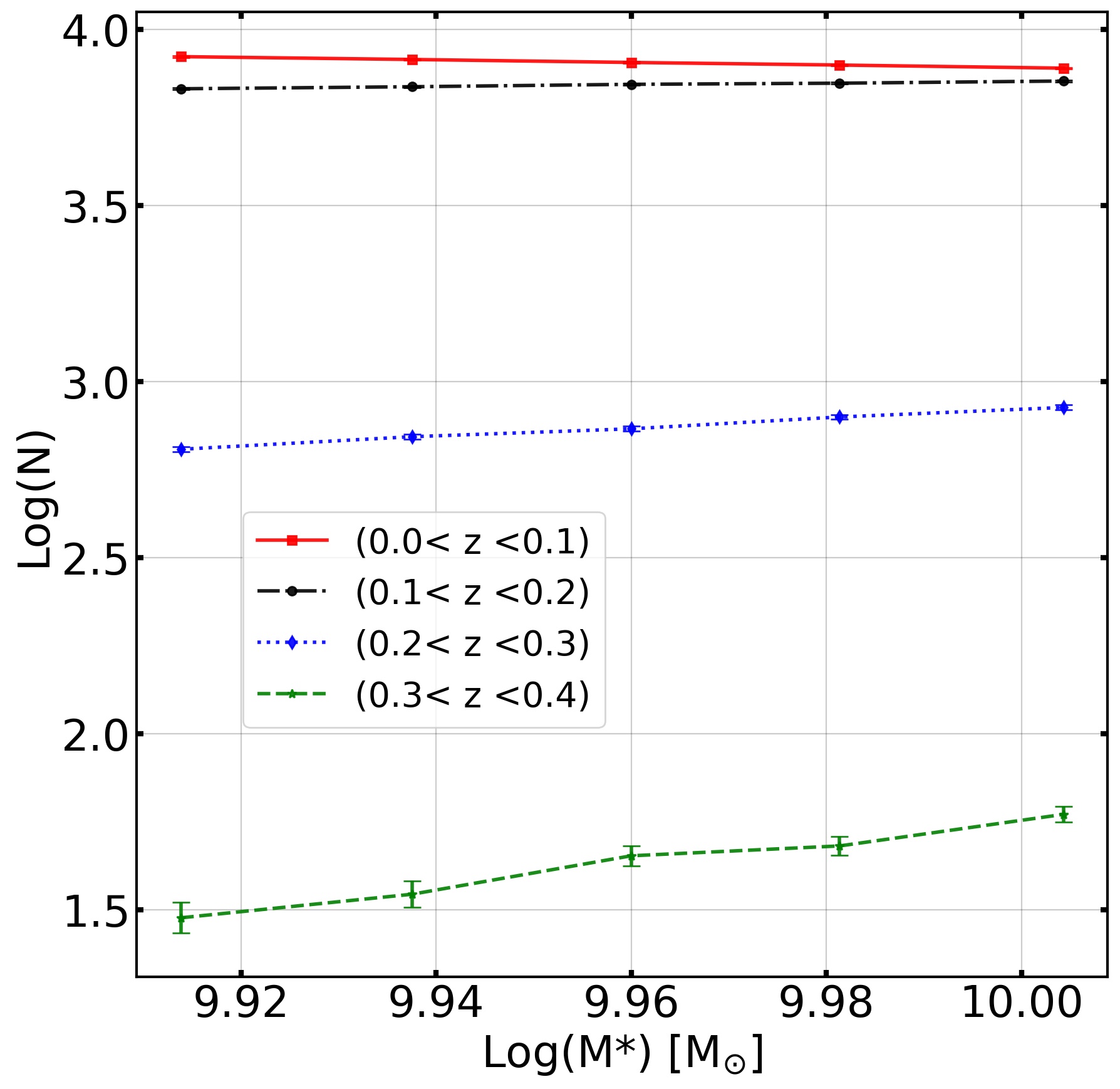}
		\centering\includegraphics[width=.85\columnwidth]{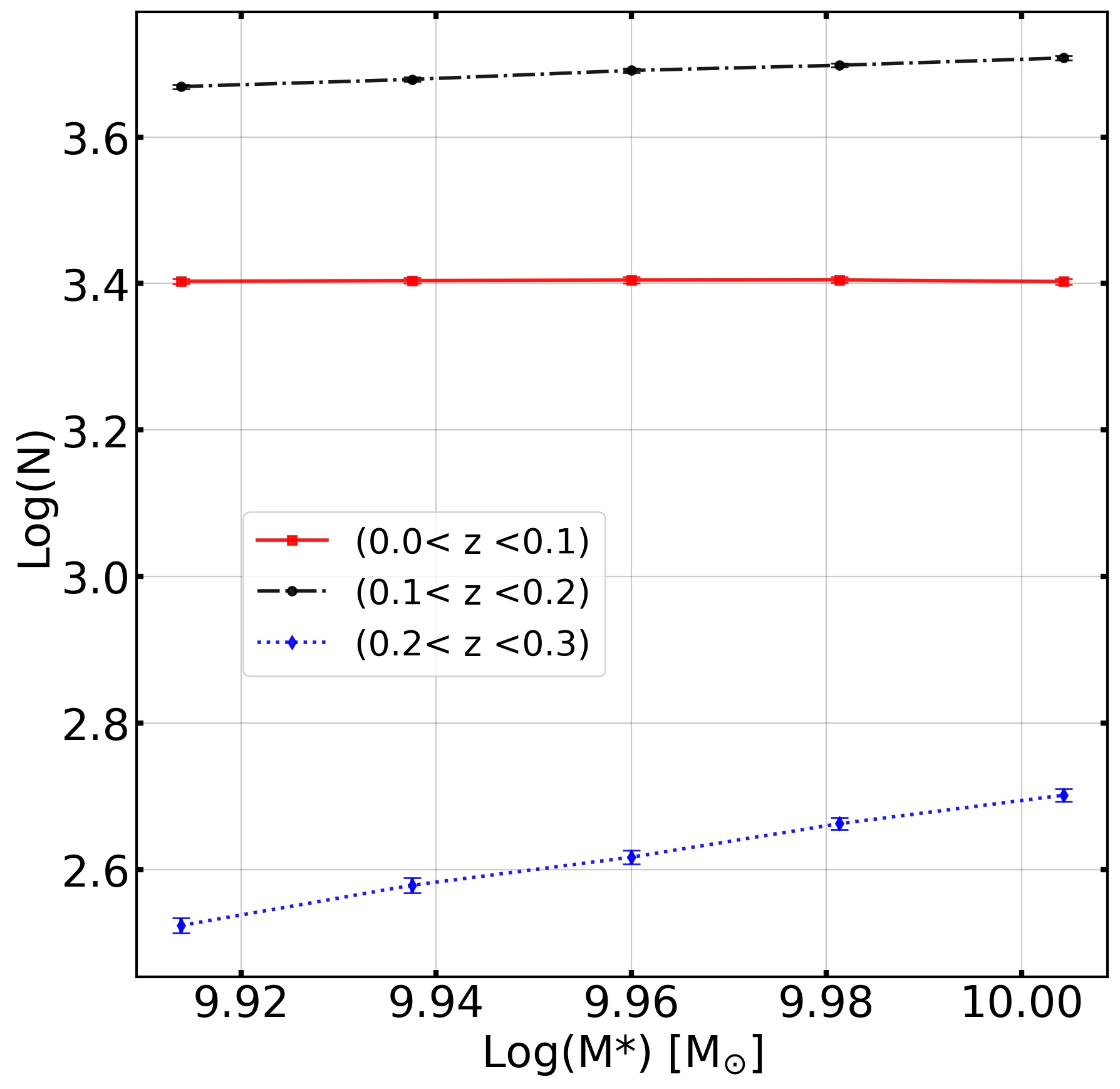}
		\caption{The sensitivity results of MIGHTEE-HI: The left panel shows the variation in the number counts with knee mass $M^*$. In the right panel, $W_{20}>200$ km/s cut is taken that results in a much steeper slope compared to the latter panel in the $0.2<z<0.3$ bin. The middle value on the x-axis represents the ALFALFA value of the knee mass.}\label{fig7}
	\end{figure*}
	
	\begin{figure*}[hbt]
		\centering\includegraphics[width=.85\columnwidth]{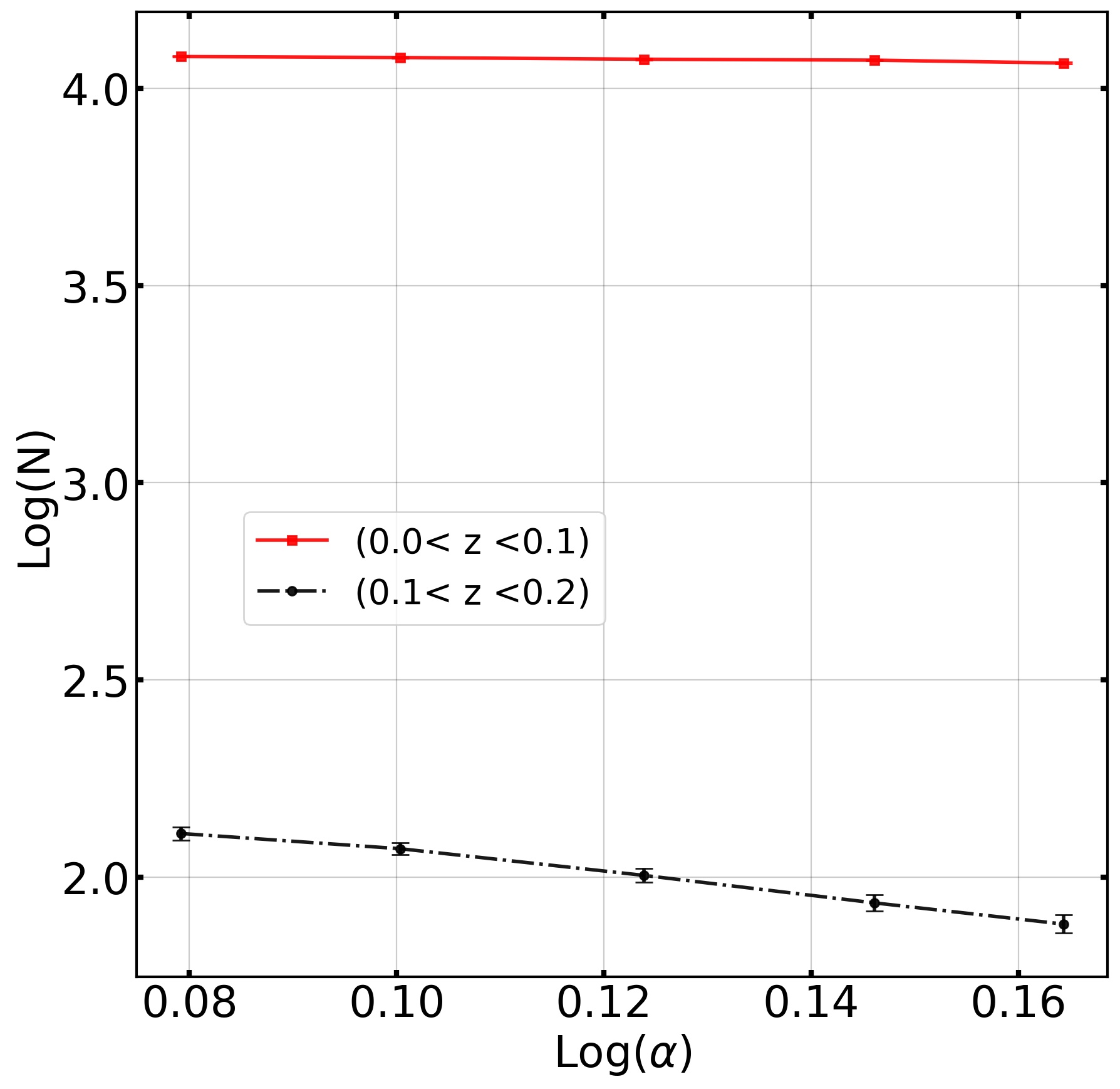}
		\centering\includegraphics[width=.85\columnwidth]{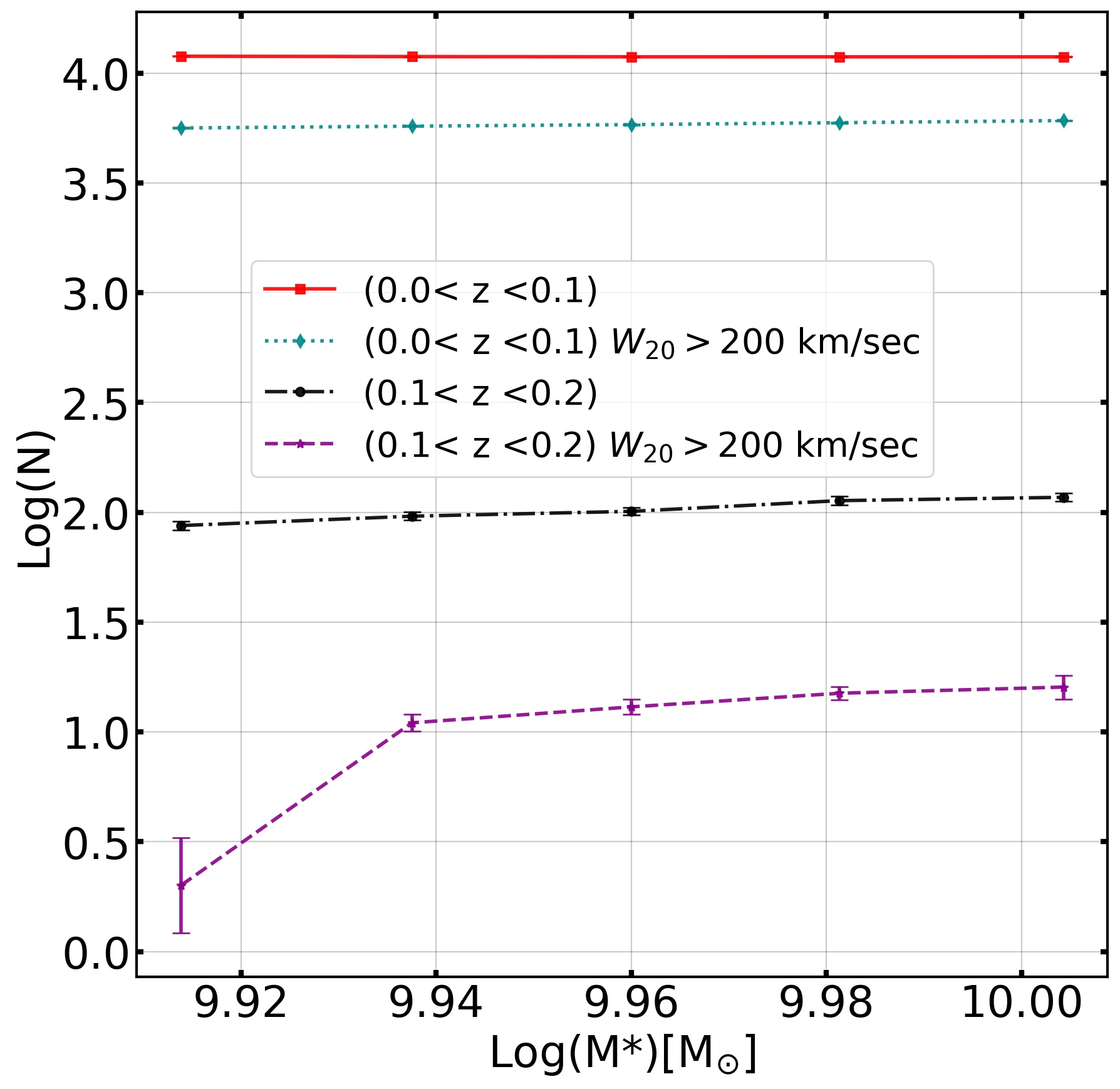}
		\caption{Sensitivity results from WALLABY: The left panel represents the number variation with the absolute value of the low-mass end slope $\alpha$. The right panel shows the sensitivity of the knee mass $M^*$ to the number without cut and a $W_{20}>200$ km/s cut. These results taken from $15$ pointings of the WALLABY survey at $5$-$\sigma$ SNR, integration time $8$ h / pointing.}\label{fig8}
	\end{figure*}
	The predicted number of detections can be affected if the survey volume is not representative, e.g., if it is dominated by a large structure.  
	Note that the Fornax cluster is included in the planned observation fields for the MIGHTEE-HI survey.  
	MeerKAT will observe this cluster as a part of the MIGHTEE-HI survey, with sky coverage of $\approx12$ deg$^2$. The Fornax cluster survey will help to improve the understanding about galaxy formation and evolution, as it is the site where most galaxy mergers and evolution take place \citep{43}.

	We take the ALFALFA HI mass function and the proposed parameters for WALLABY survey given in the table \ref{Tab3} to estimate {\it rms} noise and flux density. 
	We take $15$ (with $8$ hrs of integration time each) independent ASKAP pointings each corresponding to $\approx 30$ deg$^2$ area in our simulation. 
	It is estimated that nearly $748$ galaxies per pointing with $5$-$\sigma$ confidence can be detected. 
	The distribution of the number of detected galaxies as a function of redshift and HI masses is shown in the figure \ref{fig5}. Here, we simulated a smaller area of the sky approximately $30$ deg$^2$ that is equivalent to single field (i.e. one pointing). The number count in figure \ref{fig5} represents the galaxies under $15$ such pointings. Scaling of number counts according to the full WALLABY survey area $30940$ deg$^2$ gives nearly $7.4\times 10^5$ galaxies with $5$-$\sigma$.
	We adopt Model-III for these estimates of number counts. ASKAP can be expected to observe HI emission from almost half a million nearby galaxies in the full WALLABY survey. 
	The number of galaxies as a function of SNR from the mock catalog is presented in the right panel of the figure \ref{fig4}. 
	There is no galaxy detected within $(0.2<z<0.26)$ with SNR$\geq 3$.  
	Thus, we expect that the survey will be effectively limited to $z \leq 0.2$.  
	Although the survey is shallow compared to MIGHTEE-HI, it covers a large part of the sky and hence is expected to be representative.  
	The survey can also be used to study environmental effects, e.g., a recent pilot survey of WALLABY \citep{44} has studied the gas dynamics within Hydra I cluster. 
	Some early science observations \citep{45}, \citep{46} also suggest that ASKAP can resolve a large number of low HI mass clouds to study the tidal interaction of satellite galaxies.


	As discussed before, our number counts are validated with previous work if we adjust each parameter according to \cite{20} for the MIGHTEE-HI survey. For a representative galaxy profile $150$ km/s wide, channel width $\Delta\nu=26$ kHz, and HI mass function from \cite{26}, mass-independent size $D_{HI}=50$ kpc, we estimate {\it rms} noise $\sigma_{rms}$ and signal $S_v$. Our estimate is shown in the third column of the table \ref{Tab2} along with the comparison to the previous work (second column). As assumptions are made therein, the effect of primary beam, inclination, mass-dependent circular velocity, mass-dependent size, and cosmic variance is ignored, and so we did the same to validate the number counts. Though our numbers are a little higher, this could be the distributed noise that we used instead of a fixed flux cut. These number counts are conservative to the above assumptions and therefore underestimated. The number counts becomes $\approx 5$ times higher if we take all the factors into account, as our Model III suggests; see the left panel of figure \ref{fig4}. Our Model III uses different HIMF measured using the 2DSWML method in \cite{1}. This HIMF also makes the number counts higher than what is used from \cite{26}.
	
	The WALLABY predictions made in \citealt{22} used the fixed computed noise $1\sigma_{rms}= 1.592$ mJy km/s. They also made use of mass-dependent circular velocity, inclination, and mass-dependent size similar to our Model III. The primary beam response has been ignored in their work. Our estimate of number counts in the validation exercise is higher than theirs by $2\%$; see table \ref{Tab2}. This may be due to a different cosmology used in their model \{$\Omega_m$, $\Omega_{\Lambda}$, $h$, $\Omega_{b}$, $\sigma_8$\} = \{0.25, 0.75, 0.73, 0.045, 0.9\}.
	
	We also investigate the sensitivity of predicted number counts to parameters of the HIMF. We vary one parameter at a time and keep others fixed to the fiducial ALFALFA values. 
	For example, as we vary $\alpha$ or $M^*$, $\Omega_{HI}$ is kept fixed. To ensure this, we vary $\phi^*$ by an appropriate amount. We keep this fixed as the observations \citep{47}, \citep{48} do not show any indication of the evolution with redshift in $\Omega_{HI}$. However, the error bars are quite large in these observations. To account for the variation of $\Omega_{HI}$ within the error bars allowed up to $z\leq 0.4$, we separately vary the HI density parameter (see right panel of figure \ref{fig6} and table \ref{Tab4}) to check the effect on the number counts. We find that the number counts increase slightly, but they are not a strong function of the density parameter or the redshift.
	
	The sensitivity of the number counts for these parameters for the MIGHTEE-HI survey is shown in figures \ref{fig6} and \ref{fig7}. 
	Sensitivity can be quantified by the slope of the logarithmic derivative of number counts with respect to the HIMF parameters. 
	These relations and slopes are given in table \ref{Tab4}.
	Slopes and error bars are derived by doing a least square fit over the five data points for variation. 
	Our number counts are very sensitive to the low mass end slope $\alpha$ of the HI mass function.
	A point to note is that the slope changes sign from the first redshift bin to later bins.  
	A similar pattern is seen for the evolution of $M^*$, though the change in number counts is much smaller in this case.  
	In the right panel of figure \ref{fig6}, the change in number counts with the density parameter $\Omega_{HI}$  remains almost fixed with the redshift, except for the last redshift bin. 
	We also analyze the sensitivity of these parameters for conditional number counts: we impose cuts in the observables $W_{20}$ and $M_{HI}$.
	These are also presented in the table. 
	The main conclusion from this analysis is that it should be easy to detect any changes in $\alpha$ from these surveys, as the variation of the number counts alone is enough for this purpose.  
	The counts are less sensitive to $M^*$ as compared to $\alpha$. 
	At a fine level, one also has to consider a potential degeneracy between variations. 
	\begin{figure*}[!hbt]
		\centering\includegraphics[width=.85\columnwidth]{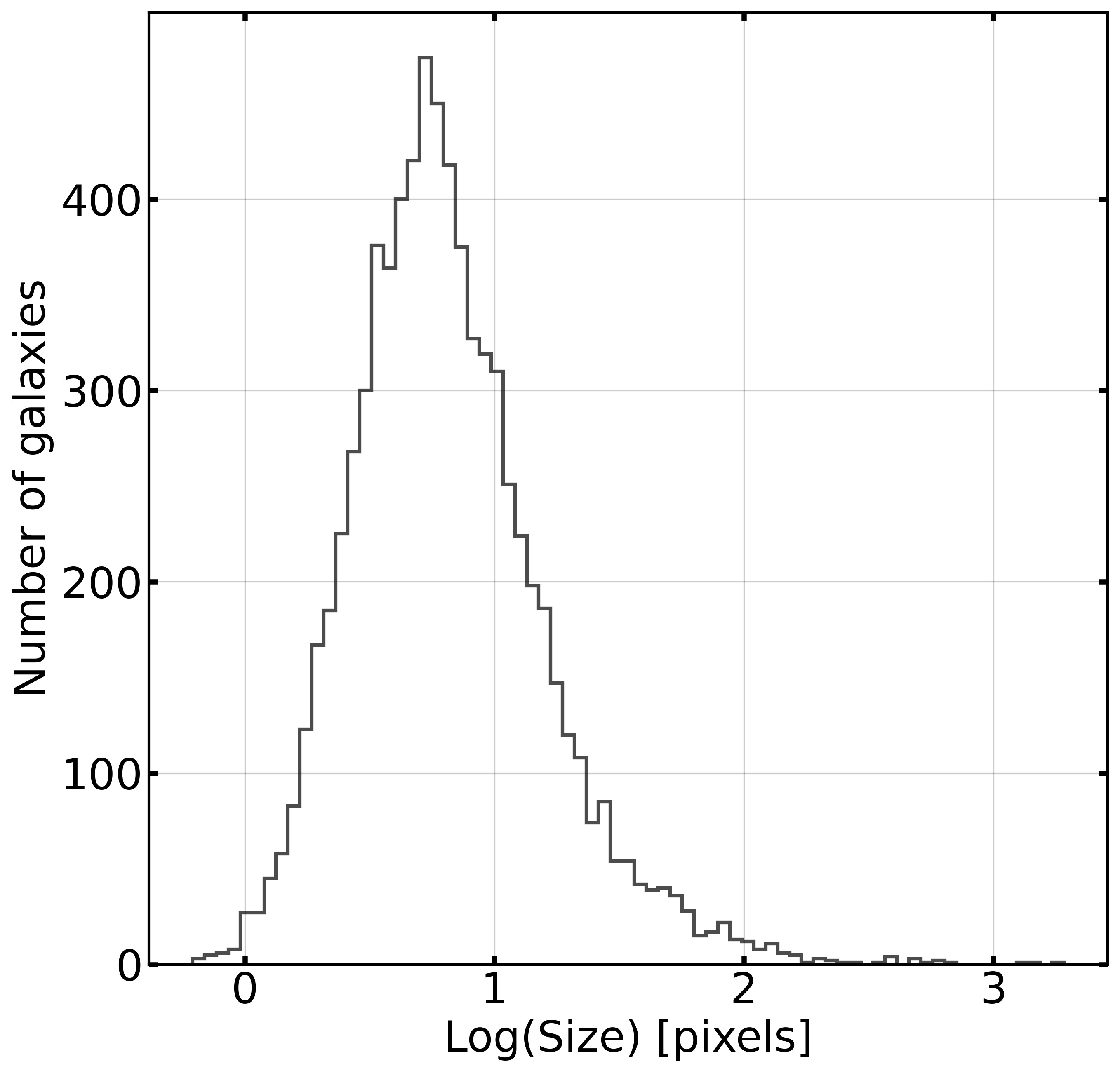}
		\centering\includegraphics[width=.85\columnwidth]{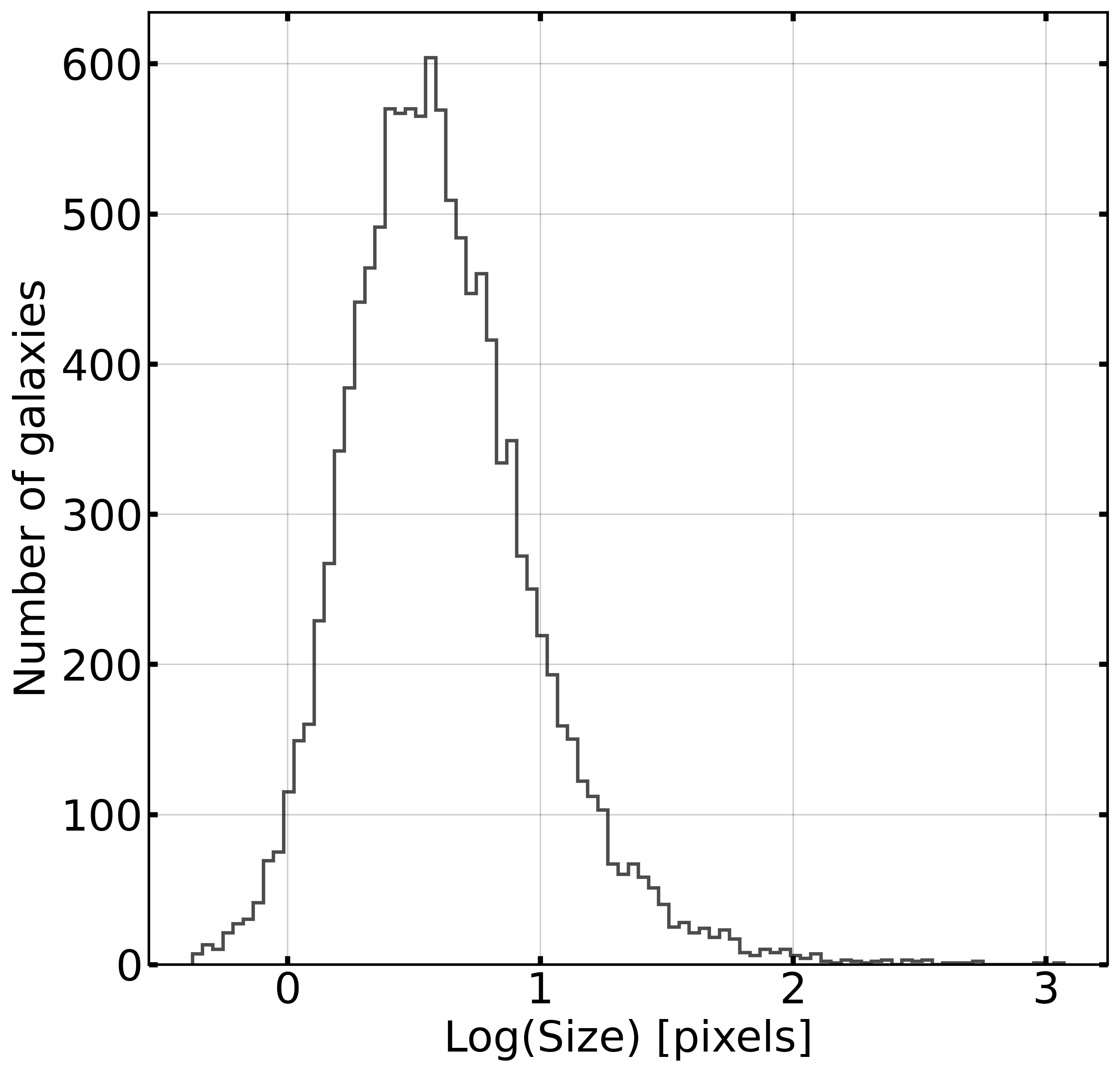}
		\caption{The distribution of the galaxy size detected blindly with $5$-$\sigma$ using model III: In the left panel, MIGHTEE-HI galaxies are shown with native pixel size $8''$ and most galaxies have a size of $2.5$ to $10$ pixels. The right panel represents the size distribution of galaxies in the WALLABY survey with native pixel size $30''$. Here, most galaxies have size of $2$ to $6$ pixels. In case of WALLABY, these numbers must be scaled with a factor of $68$ as these numbers only represent the result of $15$ individual pointings.}\label{fig9}
	\end{figure*}
	
	\begin{figure*}[!hbt]
		\centering\includegraphics[width=.85\columnwidth]{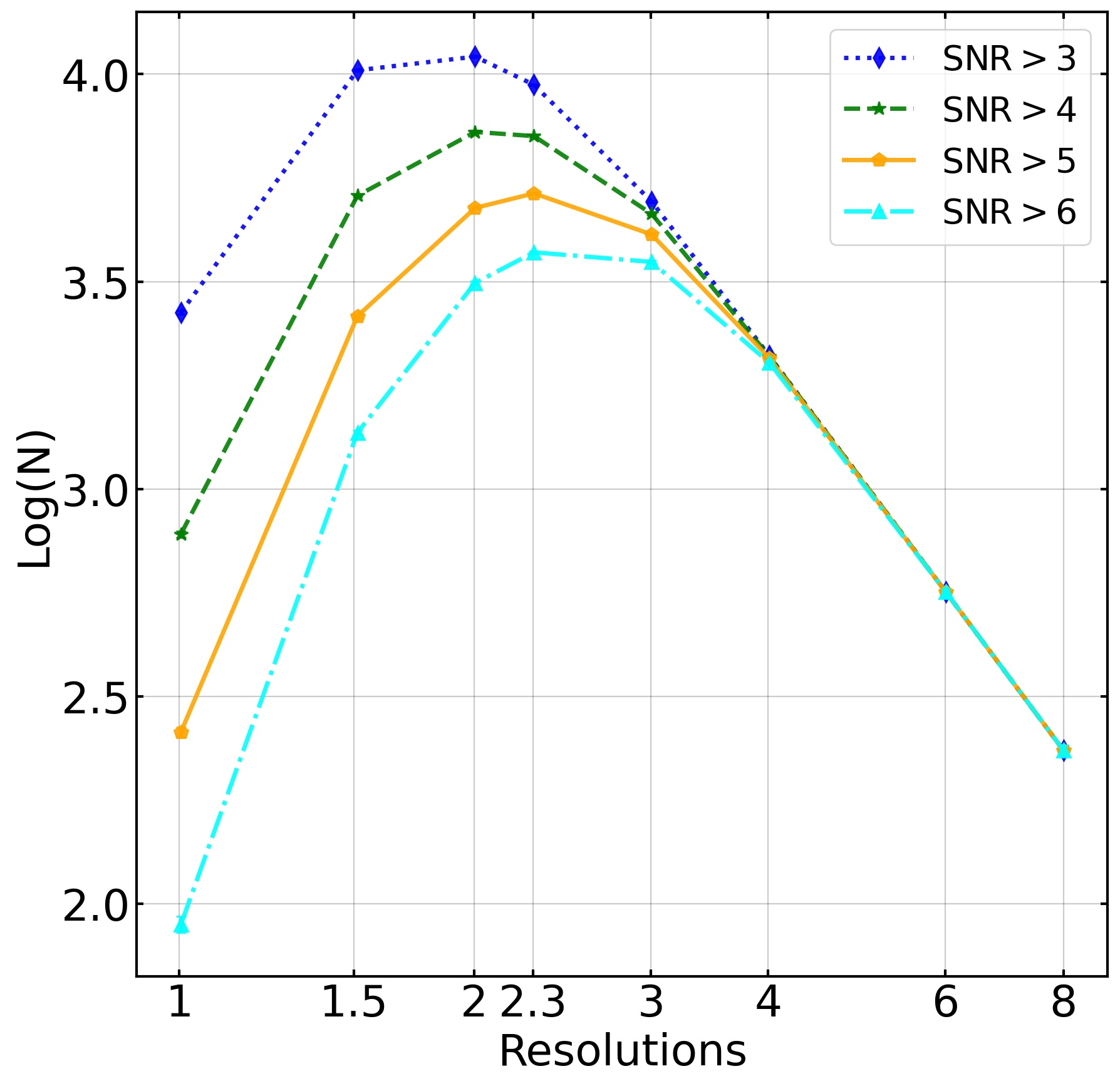}
		\centering\includegraphics[width=.85\columnwidth]{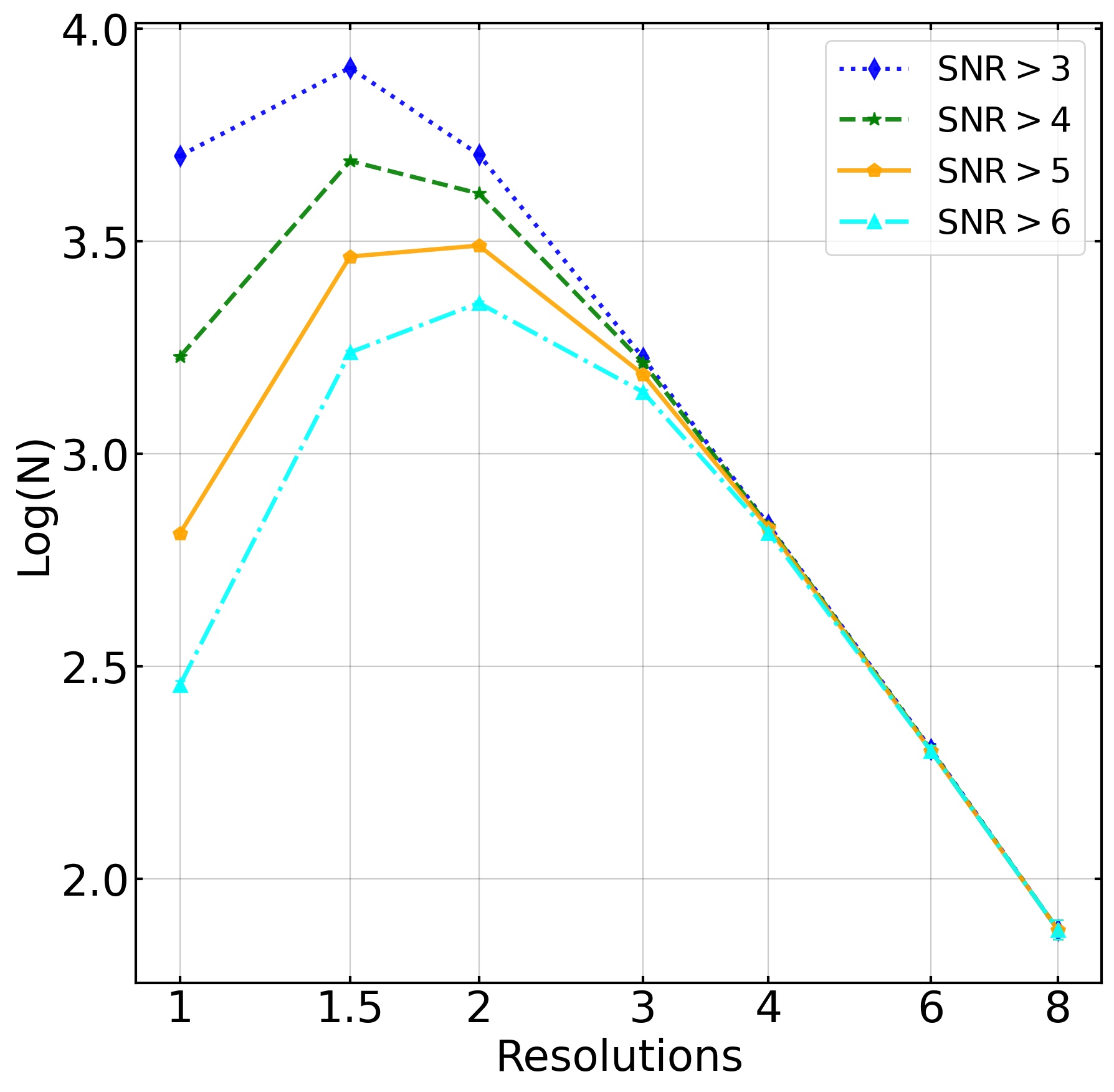}
		\caption{The number counts on the map at different resolutions using Model III: The left panel shows the logarithmic number counts at different resolutions in the MIGHTEE-HI survey map. The maximum number of galaxies is detected with $5$-$\sigma$ in the map if the images are made at $2.3$ resolution in the full MIGHTEE-HI survey. The right panel represents the same number counts in the WALLABY survey in the reduced sky area (i.e. $15$ individual pointings). The number counts are maximum when the images are made at $2$ pixel resolution. Here, the number counts must be scaled with the scale factor of $68$ to obtain the total number in the WALLABY survey.}\label{fig10}
	\end{figure*}
	The sensitivity of number counts to these parameters is also explored with the different HI masses and velocity-width cuts. 
	As one can see in the right panel of figure \ref{fig7}, the knee mass $M^*$ is sensitive to the numbers at $W_{20} > 200$ km/s in the $0.2<z<0.3$ bin than without any cuts.
	The sensitivity of the HIMF parameters for different cuts is given in tables \ref{Tab4} and \ref{Tab5}. 
	The variation of number counts expressed in terms of the logarithmic derivative with respect to the parameter and its variation with the redshift is a good indicator of the sensitivity for a specific parameter. 
	One can notice this variation for the low mass end slope $\alpha$ and knee mass $M^*$.
	The estimated error in determination of the slope is also mentioned in the table. 
	Clearly, situations where the slope is much larger than the estimated error are promising probes of the HIMF. 
	
	We have not included any variation with redshift of the HIMF parameters in our simulation. 
	This aspect of the parameters will be explored in future work. 
	Given that we already have estimates of $\Omega_{HI}$ in this redshift range that indicates little or no variation \citep{47}, \citep{48}, \citep{15}, we can guess what to expect from the variation of the number counts with the individual parameters given here. 
	The sensitivity results of the parameters for the WALLABY survey are shown in figure \ref{fig8}. 
	We do not present the HI density sensitivity figure for the WALLABY survey, but the corresponding sensitivities for the parameters can be seen in the table \ref{Tab5}. 
	We find that the number counts are very sensitive to $\alpha$ for the WALLABY survey as well, in spite of somewhat different survey specifications.

	We would like to comment on the dependence of the number counts on the method chosen for computing SNR for a given galaxy in the mock catalog.  
	We have assumed that the resolution can be continuously varied to optimize the SNR.  
	This approach can be computationally expensive.  Here we provide results for an approach where the maps are analyzed at a few fixed resolutions for blind detection.  
	A factor of relevance for blind detection is that most galaxies have approx and angular extent that is $2.5$ to $10$ pixels size range, see the left panel of figure \ref{fig9} for MIGHTEE-HI survey. In the WALLABY survey, the pixels range where most of the galaxies lie is $2$ to $6$; see the right panel of figure \ref{fig9}. WALLABY numbers counts given here are for $15$ observed fields and need to be scaled for full survey. MeerKAT and ASKAP have a pixel size (i.e., 1 resolution) equal to $8$ arcsec and $30$ arcsec, respectively.

	In synthesis imaging, the sensitivity of the flux density depends on $(1)$ the number of baselines and $(2)$ baseline distribution. This information is used to construct the maps. The SNR depends on the number of baselines but not on their distribution, but the resolution of the image depends on the baseline distribution. If the galaxy is resolved, the largest baselines have little or no correlated flux-density contribution. Therefore, using the longer baselines in the synthesis imaging only adds their noise to the map but no information about the signal from the galaxy. Only those baselines that contribute to the signal can enhance the detectability of the galaxy. We optimize the SNR of a given galaxy size in the map by making the images at different resolution so that it may get detected at one or the other resolution. The number of galaxies counted at different resolutions is presented in the figure \ref{fig10}. The left panel represents the number counts for the MIGHTEE-HI survey with $1$- resolution scale equal to $\approx8''$. The maximum number of galaxy are detected at $2.3$ resolutions in the MIGHTEE-HI survey. At coarser resolutions (i.e. above the $1$ resolution in the figure \ref{fig10}), the noise in the image is greater than in the image at finer resolution. At coarser resolution, the given size of a galaxy is concentrated into a small number of resolution elements (i.e. pixels), raising the flux density per resolution element faster than noise. Therefore, at first the SNR per pixel improved on coarser resolution, giving rise to an increasing number count up to $2.3$ pixel resolution. At much coarser resolution (beyond $2.3$), {\it rms} noise dominates the image, leading to loss of the number counts of SNR $>5$ per pixel. The right panel shows the number counts for the WALLABY survey with $1$-resolution scale equal to $\approx30''$.

	\section{Summary}
	
	The next generation radio instruments are coming online and these are expected to carry out surveys with an unprecedented sensitivity. 
	SKA precursors like MeerKAT and ASKAP are amongst the upcoming telescopes that can carry out large-scale surveys in the redshifted $21$~cm line and observe emission from atomic hydrogen in galaxies.  
	Mock catalogs and pilot surveys are used to optimize survey strategies to maximize the scientific returns of such programs.  
	With very large-scale projects, pre-pilot surveys that mimic the planned survey strategy give us important insight into instrumental performance and sensitivity \citep{49} studies. 
	
	In this paper, we have presented our simulation method for creating mock catalogs for large-scale surveys of galaxies in the redshifted $21$~cm line of atomic hydrogen.  
	Our predictions show that MeerKAT may detect $\approx 15800$ ($525$ galaxies per pointing) in MIGHTEE-HI survey with SNR $> 5$ to redshift $z=0.4$. 
	ASKAP may detect nearly $7.5\times 10^5$ galaxies in a $30940$ deg$^2$ area of the sky as proposed in the WALLABY survey. With the help of the depth and numbers covered by these surveys, we can make progress in studying HIMF, the 2-point correlation function \citep{50}, and the kinematics of the gas. 
	Commensal observations in the continuum will permit us to estimate star formation rates in these galaxies.
	
	Our key findings are as follows:
	\begin{itemize}
		\item 
		MIGHTEE-HI will do $30$ pointings and is capable of blind detection of galaxies up to $z=0.4$.  We find that the number counts are sensitive to the HIMF and hence it should be possible to constrain the HIMF parameters very well.  Further, given the redshift extent, it may be possible to constrain any evolution, especially the low mass index $\alpha$.
		\item 
		Our predictions for MIGHTEE-HI are much larger than those of earlier estimates.  We believe that this is due to two reasons: one is that we take more factors into account and the incorporation of random orientations and mass-dependent size leads to an enhancement as soon in comparison of Models I and II.  The other reason is that we have assumed independent pointings, whereas MIGHTEE-HI will have overlap between fields.
		\item 
		The WALLABY survey covers a large fraction of the sky, though it is not as deep as the MIGHTEE-HI survey.  The expected number counts for this survey are close to one million.  We find that there will be few galaxies detected at redshifts beyond $0.2$. 
		\item 
		We show that even gross number counts, and subsets with cuts based on observed quantities can be used to constrain HIMF parameters in both the surveys. 
		\item 
		The optimal range to build maps with angular resolution is $[2,4]$ pixels for the MIGHTEE-HI survey and $[1.5,3]$ pixels for the WALLABY survey.  This permits detection of a vast majority of galaxies with considerably smaller computational cost.  Of course, it may be worthwhile to explore further optimization of peaks in signal with a slightly lower SNR, say 4, in order to detect some more galaxies.
		
	\end{itemize}
	
	We need to improve our predictions for WALLABY by accounting for the large range of declinations that it is expected to observe. 
	
	We plan to improve our simulations by adding more attributes of galaxies: radio continuum, optical properties, etc. 
	This will permit us to make cuts based on selection in different bands and also to make estimates based on joint selections between radio and optical surveys.
	Lastly, we can combine an approach based on the halo model and N-Body simulations to include information about clustering and environment.  
	We expect to include these in the simulation in the coming months. 
	
	\section*{Acknowledgements}
	
	The authors acknowledge the use of IISER Mohali HPC
	facility. This research has used NASA’s Astrophysics Data System Bibliographic Services. The author thanks the Department of Science and Technology (DST), Government of India, for financial support through Council of Scientific and Industrial Research-UGC research fellowship. The authors would like to thank the anonymous referee for the helpful comments.
	
	\bibliographystyle{apj}
	\bibliography{manuscript}

\end{document}